\documentclass[twocolumn]{aastex63}

\hypersetup{linkcolor=blue,citecolor=blue,filecolor=cyan,urlcolor=blue}

\usepackage{CJK}

\usepackage{longtable}
\usepackage{lipsum} 
\usepackage{multirow}

\usepackage{natbib}
\usepackage{graphicx}
\usepackage{bm}
\usepackage{amsmath}
\usepackage{amsfonts}
\usepackage{amssymb}
\usepackage{xcolor}
\usepackage{mathrsfs}
\usepackage[caption=false]{subfig}
\usepackage{CJK}
\usepackage{ulem}

\usepackage{booktabs}

\shorttitle{Chemistry}
\shortauthors{Li et al.}
\usepackage{CJK}

\graphicspath{{./}{figures/}}
\usepackage{xspace}
\newcommand{\um}{{$\mu$m}\xspace}
\def\arcsec{$^{\prime\prime}$\xspace}

\def\kms{km~s$^{-1}$\xspace}
\def\c18o{C$^{18}$O}
\def\dcop{DCO$^{+}$\xspace}
\def\n2dp{N$_{2}$D$^{+}$}
\def\ccd{C$_{2}$D\xspace}
\def\ch3oh{CH$_{3}$OH}
\def\h2co{H$_{2}$CO}
\def\1{\uppercase\expandafter{\romannumeral1}}
\def\2{\uppercase\expandafter{\romannumeral2}}
\def\3{\uppercase\expandafter{\romannumeral3}}
\def\4{\uppercase\expandafter{\romannumeral4}}
\def\5{\uppercase\expandafter{\romannumeral5}}
\def\6{\uppercase\expandafter{\romannumeral6}}
\def\7{\uppercase\expandafter{\romannumeral7}}

\newcommand{\x}{\textcolor{red}{XX}}

\begin{document}

\title{The ALMA Survey of 70 $\mu \rm m$ Dark High-mass Clumps in Early Stages (ASHES). \7: Chemistry of Embedded Dense Cores}

\correspondingauthor{Shanghuo Li}
\email{shanghuo.li@gmail.com}

\author[0000-0003-1275-5251]{Shanghuo Li }
\affiliation{Korea Astronomy and Space Science Institute, 776 Daedeokdae-ro, Yuseong-gu, Daejeon 34055, Republic of Korea}
\affiliation{Max Planck Institute for Astronomy, Königstuhl 17, D-69117 Heidelberg, Germany}

\author[0000-0002-7125-7685]{Patricio Sanhueza}
\affiliation{National Astronomical Observatory of Japan, National Institutes of Natural Sciences, 2-21-1 Osawa, Mitaka, Tokyo 181-8588, Japan}
\affiliation{Department of Astronomical Science, SOKENDAI (The Graduate University for Advanced Studies), 2-21-1 Osawa, Mitaka, Tokyo 181-8588, Japan}

\author[0000-0003-2619-9305]{Xing Lu}
\affiliation{Shanghai Astronomical Observatory, Chinese Academy of Sciences, 80 Nandan Road, Shanghai 200030, People's Republic of China} 

\author[0000-0002-3179-6334]{Chang Won Lee}
\affiliation{Korea Astronomy and Space Science Institute, 776 Daedeokdae-ro, Yuseong-gu, Daejeon 34055, Republic of Korea}
\affiliation{University of Science and Technology, 217 Gajeong-ro, Yuseong-gu, Daejeon 34113, Republic of Korea}

\author[0000-0003-2384-6589]{Qizhou Zhang}
\affiliation{Center for Astrophysics $|$ Harvard \& Smithsonian, 60 Garden Street, Cambridge, MA 02138, USA}

\author[0000-0003-2814-6688]{Stefano Bovino}
\affiliation{Departamento de Astronom\'ia, Facultad Ciencias F\'isicas y Matem\'aticas, Universidad de Concepci\'on, Av. Esteban Iturra s/n Barrio Universitario, Casilla 160, Concepci\'on, Chile}

\author[0000-0002-6428-9806]{Giovanni Sabatini}
\affiliation{INAF - Istituto di Radioastronomia - Italian node of the ALMA Regional Centre (It-ARC), Via Gobetti 101, I-40129 Bologna, Italy}

\author[0000-0002-5286-2564]{Tie Liu}
\affiliation{Shanghai Astronomical Observatory, Chinese Academy of Sciences, 80 Nandan Road, Shanghai 200030, People's Republic of China}

\author[0000-0003-2412-7092]{Kee-Tae Kim}
\affiliation{Korea Astronomy and Space Science Institute, 776 Daedeokdae-ro, Yuseong-gu, Daejeon 34055, Republic of Korea}
\affiliation{University of Science and Technology, 217 Gajeong-ro, Yuseong-gu, Daejeon 34113, Republic of Korea}

\author[0000-0002-6752-6061]{Kaho Morii}
\affiliation{Department of Astronomy, Graduate School of Science, The University of Tokyo, 2-21-1, Osawa, Mitaka, Tokyo 181-0015, Japan}

\author[0000-0002-2149-2660]{Daniel Tafoya}
\affiliation{Department of Space, Earth and Environment, Chalmers University of Technology, Onsala Space Observatory, 439~92 Onsala, Sweden}

\author[0000-0002-8149-8546]{Ken'ichi Tatematsu}
\affiliation{National Astronomical Observatory of Japan, National Institutes of Natural Sciences, 2-21-1 Osawa, Mitaka, Tokyo 181-8588, Japan}

\author[0000-0003-4521-7492]{Takeshi Sakai}
\affiliation{Graduate School of Informatics and Engineering, The University of Electro-Communications, Chofu, Tokyo 182-8585, Japan.}

\author[0000-0001-6106-1171]{Junzhi Wang}
\affiliation{Shanghai Astronomical Observatory, Chinese Academy of Sciences, 80 Nandan Road, Shanghai 200030, People's Republic of China} 

\author[0000-0002-9832-8295]{Fei Li} 
\affiliation{School of Astronomy and Space Science, Nanjing University, 163 Xianlin Avenue, Nanjing 210023, People's Republic of China}

\author[0000-0001-9500-604X]{Andrea Silva}
\affiliation{National Astronomical Observatory of Japan, National Institutes of Natural Sciences, 2-21-1 Osawa, Mitaka, Tokyo 181-8588, Japan}

\author[0000-0003-1604-9127]{Natsuko Izumi}
\affiliation{Academia Sinica Institute of Astronomy and Astrophysics, 11F of AS/NTU Astronomy-Mathematics Building, No.1, Section 4, Roosevelt Road, Taipei 10617, Taiwan}

\author[0000-0002-4173-2852]{David Allingham}
\affiliation{School of Mathematical and Physical Sciences, University of Newcastle, University Drive, Callaghan, NSW 2308, Australia}




\begin{abstract}
We present a study of chemistry toward 
294 dense cores in 12 molecular clumps using the data 
obtained from the ALMA Survey of 70 $\mu \rm m$ dark 
High-mass clumps in Early Stages (ASHES). We identified  
97 protostellar cores and 197 prestellar 
core candidates based on the detection of outflows and molecular 
transitions of high upper energy levels ($E_{u}/k > 45$ K).  
The detection rate of the N$_{2}$D$^{+}$ emission toward the 
protostellar cores is 38\%, which is higher than 9\% for 
the prestellar cores, indicating that N$_{2}$D$^{+}$ does not 
exclusively trace prestellar cores. The detection rates of the 
DCO$^{+}$ emission are 35\% for the prestellar cores and 49\% 
for the protostellar cores, which are higher than those  of 
N$_{2}$D$^{+}$, implying that DCO$^{+}$ appears more 
frequently than N$_{2}$D$^{+}$ in both prestellar and protostellar 
cores. Both N$_{2}$D$^{+}$ and DCO$^{+}$ abundances appear 
to  decrease from the prestellar to protostellar stage.   
The DCN, C$_{2}$D and $^{13}$CS emission lines are rarely 
seen in the dense cores of early evolutionary phases.  
The detection rate of the H$_{2}$CO emission toward dense cores 
is 52\%, three times higher than that of CH$_{3}$OH (17\%).  
In addition, the H$_{2}$CO detection rate, abundance, 
line intensities, and line widths increase with the core 
evolutionary status,  suggesting that the H$_{2}$CO line 
emission is sensitive to protostellar activity.

\end{abstract}



\section{Introduction} 
\label{sec:intro}
The chemical composition of planets is affected 
by the chemical makeup of protoplanetary disks within 
which they form. The chemical  content of prestellar 
and protostellar cores  sets the initial conditions 
in protoplanetary disks 
\citep{2012A&ARv..20...56C,2019MNRAS.490...50D,
2020ARA&A..58..727J,2021NatAs...5..684B,
2021PhR...893....1O}. 
Molecular lines are a powerful tool to reveal the 
chemical and physical processes during star formation 
and core evolution,  since different molecules can be 
associated with specific chemical and physical 
environments \citep{2007ARA&A..45..339B}. 
Consequently, different molecular lines can be used 
to probe different gas environments, i.e., different 
physical conditions. For instance, deuterated molecules 
(e.g., \n2dp, \dcop, H$_{2}$D$^{+}$, NH$_{2}$D) can be 
used to trace cold and dense molecular clumps/cores 
associated with early evolutionary stages of star 
formation 
\citep[e.g., prestellar cores;][]{2002ApJ...565..344C,
2017ApJ...834..193K,2019A&A...621L...7G,
2020A&A...644A..34S,2021A&A...650A.202R,
2021ApJ...912L...7L,2022ApJ...925..144S}. 
On the other hand, the high density gas tracers 
could also suffer from depletion toward the cold   
and dense regions 
(e.g., N$_{2}$H$^{+}$, \citealt{2007A&A...467..179P}; 
N$_{2}$D$^{+}$, \citealt{2019A&A...629A..15R}; 
NH$_{3}$, \citealt{2022AJ....163..294P}).  
A prestellar core would further evolve into 
a protostellar core, in which protostars 
launch molecular outflows and heat the surrounding 
material.  These physical processes can make molecules 
to be released from the grain surface to the gas phase, 
causing the enhancement of various molecules in 
the gas phase \citep{1998ARA&A..36..317V,
2009ARA&A..47..427H}.  
CO and SiO are frequently used to probe 
protostellar activities \citep[e.g.,][]{2010ApJ...715...18S,
2017ApJ...841...97S,2019ApJ...878...29L,
2020ApJ...903..119L,2021ApJ...909..177L}, 
i.e., molecular jets and outflows.  
Formaldehyde (\h2co) and methanol (\ch3oh) are 
commonly seen in star-forming regions, and their 
abundances can be significantly enhanced with 
respect to quiescent regions in the presence of 
protostellar activity 
\citep[e.g., molecular outflows;][]{2008ApJ...681L..21A,
2013ApJ...773..123S,2012ApJ...754...70S,
2020ARA&A..58..727J,2021ApJ...923..147M,
2021A&A...655A..65T}. 
In addition, both species 
play a key role in the formation 
of more complex organic molecules, such as amino 
acids and other prebiotic molecules 
\citep{2002Natur.416..401B,2002Natur.416..403M,
2008ApJ...682..283G,2013A&A...560A..73G}, 
which might be transported to circumstellar disks 
and potential planetary systems \citep{2019MNRAS.490...50D}. 
Thus, a full understanding of the chemical properties 
of star-forming  clouds is essential to improving our 
knowledge  of the physical and chemical processes 
that take place during star formation.

There have been numerous  observational investigations 
aiming at understanding the chemistry of star formation 
molecular clouds. For instance, single pointing observations 
of a sample of massive clumps using single-dish telescopes 
(e.g., infrared dark clouds, IRDCs:   
\citealt{2012ApJ...756...60S,
2014ApJ...780...85V}; IRDCs to hot cores:  
 \citealt{2014A&A...563A..97G,2021A&A...652A..71S}), 
single dish mapping of a sample of massive clumps 
(e.g., IRDCs:  \citealt{2014A&A...562A...3M};  
IRDCs to HII regions:  \citealt{2013ApJ...777..157H}), 
interferometer/single-dish observations toward several 
massive clumps (e.g., IRDCs:  \citealt{2020ApJ...901..145F}; 
HII regions: \citealt{2017MNRAS.466..248L}), 
interferometer observations of a sample of HII regions 
\citep{2022MNRAS.511.3463Q}, 
and case studies  
\citep[e.g.,][]{2013ApJ...773..123S,2014A&A...572A..63I,
2020ApJ...901...31L,2022MNRAS.512.4419P}. 
Thanks to these observational studies  of chemistry 
toward different star formation regions, our understanding 
of chemistry is significantly advanced, for instance, 
chemical abundances of molecular species vary significantly 
through the evolutionary sequence of star-forming regions.

Despite these advances, the chemical properties of 
prestellar and protostellar cores are still unclear, 
due to a lack of observations of a large sample of dense 
cores at early evolutionary stages.
Here, we use ALMA high sensitivity 
and high spatial resolution data to investigate 
chemistry of a statistically significant sample 
(N = 294) of spatially resolved deeply embedded dense 
cores, which are still at extremely early evolutionary 
stages of star formation.  
The data were obtained by the ALMA Survey of 70 
$\mu \rm m$ dark High-mass clumps 
in Early Stages (ASHES) first presented in 
\citet[][hereafter Paper I]{2019ApJ...886..102S}. 
The molecular outflow content and the CO depletion fraction 
of the detected ASHES cores are presented in 
\citet[][hereafter Paper II]{2020ApJ...903..119L} and 
\cite{2022arXiv220712431S}, respectively. 
Case studies are presented in \cite{2021ApJ...913..131T}, 
\cite{2021ApJ...923..147M}, and \cite{2022ApJ...925..144S}.

Among the 294 dense cores revealed in the continuum emission, 
we have identified 197 prestellar core candidates 
(hereafter prestellar cores) and 97 protostellar cores. 
The number of prestellar and protostellar cores has been 
updated, thus is slightly different from the 210 prestellar 
and 84 protostellar cores reported in Paper~\1. 
A core is classified as prestellar (category 1) if it 
is not associated with molecular outflows and/or 
emission from any of the three lines  
CH$_{3}$OH $4_{2,2}-3_{1,2}$ ($E_{u}/k$ = 45.46 K), 
H$_{2}$CO $3_{2,2}-2_{2,1}$ ($E_{u}/k$ = 68.09 K), 
and H$_{2}$CO $3_{2,1}-2_{2,0}$ ($E_{u}/k$ = 68.11 K). 
A core is classified as an ``outflow core" (category 2) 
if it is associated with outflows 
detected in the molecular line emission   
but without  detection of any of the three 
aforementioned lines (Paper \2).   
Details of 
the molecular outflows are presented in Paper \2.  
A core is classified as a ``warm core" (category 3) if 
it is associated with emission from any of the three 
aforementioned lines but  without the outflow detection. 
The ``warm core" refers to an evolutionary stage prior 
to the ``hot core" phase. A core is classified as a 
``warm \& outflow core" (category 4) if it is 
associated with emission from any of the three 
aforementioned lines as well as with outflows. 
Overall, categories 2, 3, and 4 are considered 
as protostellar cores.

\begin{figure*}
\center
 \includegraphics[scale=0.33]{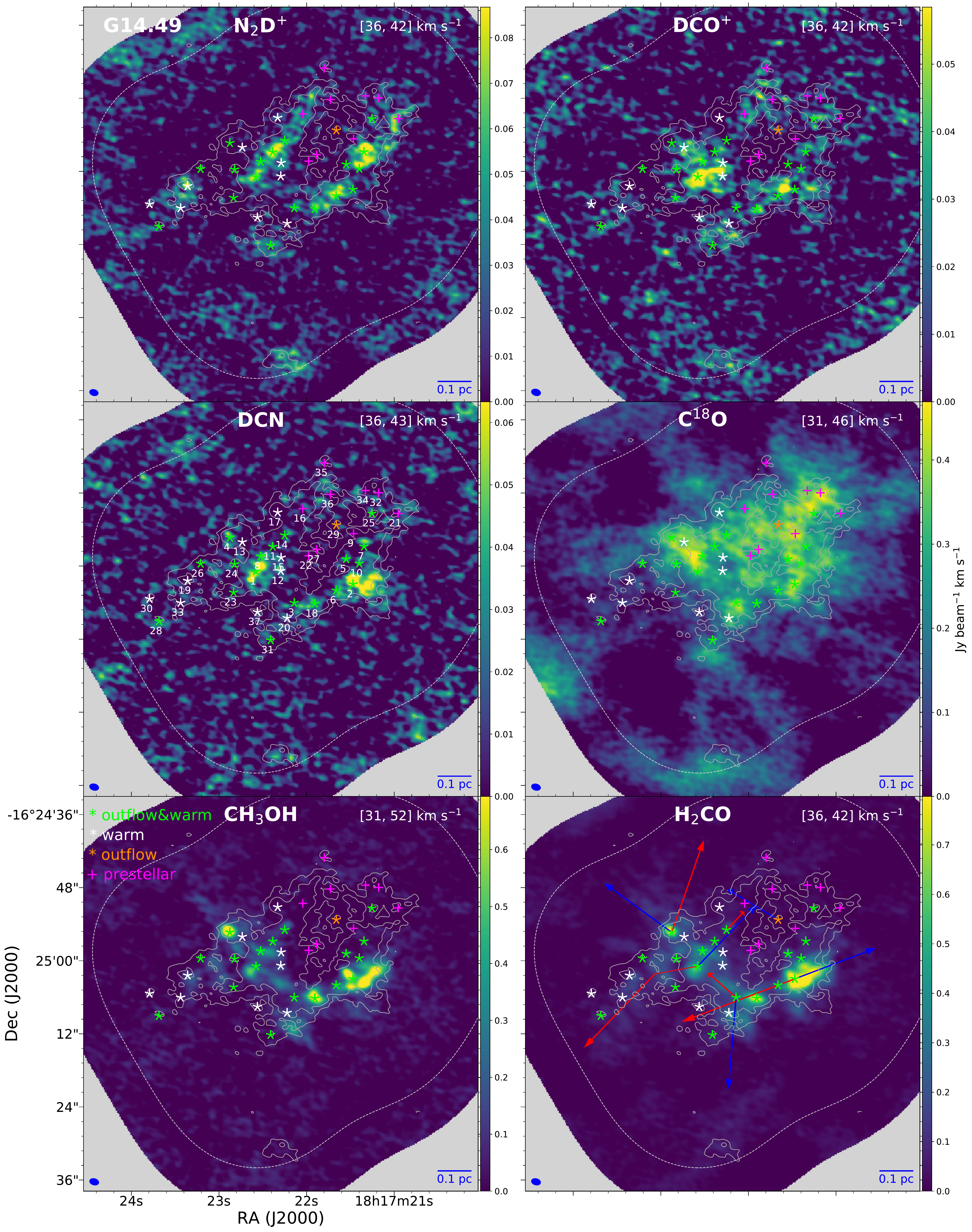}
\caption{The velocity-integrated intensity maps of \n2dp, 
\dcop, DCN, \c18o , \ch3oh, and \h2co  emission lines  
toward G14.49. The unit of the colorbar is 
Jy beam$^{-1}$ km s$^{-1}$. 
Fuchsia pluses, yellow, white, and green asterisks  
indicate prestellar candidates (category 1), outflow cores 
(category 2), warm cores (category 3), outflow+warm 
cores (category 4). 
The white dashed line shows 30\% of the sensitivity 
level of the mosaic in the ALMA continuum image. 
The blue and red arrows in bottom right panel indicate  
directions of blue-shifted and red-shifted CO outflow lobes   
(see Paper \2), respectively. 
The gray contours in each panel show the 
1.3 mm continuum emission.  The contour levels 
are (3, 6) $\times \, \sigma$, 
with $\sigma$ = 0.115 mJy beam$^{-1}$. 
The beam size of line emission and 
scale bar are shown in the lower 
left and right corner of each panel, respectively. 
The remaining sources are presented in the 
Appendix~\ref{app:fig} (see Figures~\ref{fig:comb}-\ref{fig:comb4}). 
}
\label{fig:mom0}
\end{figure*}

In this work, we study the chemistry of the embedded 
dense cores in the extremely early evolutionary stages 
of high-mass star-forming regions using high angular 
resolution and high sensitivity ALMA observations.  
With a statistically significant sample of dense cores, 
we will study the properties of deuterated molecules 
(\n2dp, \dcop, \ccd, and DCN) and dense gas tracers 
(\c18o, \h2co,  \ch3oh, and $^{13}$CS). 
The paper is organized as follows: Section~\ref{sec:obs} 
describes the observations. The results and analysis 
are  presented in Section~\ref{sec:results}. 
In Section~\ref{sec:discu} we discuss the results. 
A summary of the main conclusions is presented in 
Section~\ref{sec:conclu}.

\section{Observations} 
\label{sec:obs}
Observations of twelve 70 $\mu$m-dark molecular clumps were 
performed with ALMA in Band 6 ($\sim$224~GHz; 1.34~mm) 
using the main 12~m array, the 7~m array, and the 
total power array (TP; Project ID: 2015.1.01539.S, 
PI: P. Sanhueza). 
The mosaic observations were carried out with the 12~m 
array and 7~m array to cover a significant portion of the 
clumps, as defined by single-dish continuum images.  
The same correlator setup was applied for all sources. 
More details on the observations can be found in 
Papers I and II.

Data calibration was performed using the CASA software  
package version 4.5.3, 4.6, and 4.7, 
while the 12 m and 
7 m array datasets were concatenated and imaged 
together using the CASA 5.4 \texttt{tclean} algorithm  
\citep{2007ASPC..376..127M}. 
Data cubes for lines were produced using the 
\texttt{yclean} script which automatically clean each 
map channel with custom made masks
\citep{2018ApJ...861...14C}. 
Two times $\sigma$ root mean square 
(rms) threshold is used in the imaging process. 
The continuum emission was obtained by 
averaging the line-free channels in visibility space. 
We used a multiscale clean for continuum and 
data cube, with scale values of 0, 5, 15, and 25 times 
the image pixel size of 0\arcsec.2. 
Since some sources were observed with different 
configurations, a uv-taper was used for such 
sources in order to obtain a similar synthesized beam 
of $\sim$1\farcs2 for all sources. We adopted a Briggs 
robust weighting of 0.5 and 2 for the visibilities of 
continuum and lines in the imaging process, respectively. 
This achieved an averaged 1$\sigma$ noise level of 
$\sim$0.1~mJy\,beam$^{-1}$ for 
the continuum images. For the detected molecular lines 
(N$_{2}$D$^{+}$ $J$= 3--2,  DCN $J$= 3--2, 
DCO$^{+}$ $J$= 3--2, C$_{2}$D $J$= 3--2, 
$^{13}$CS $J$= 5--4, SiO $J$= 5--4, 
C$^{18}$O $J$= 2--1, CO $J$= 2--1,  
CH$_{3}$OH $4_{2,2}-3_{1,2}$, and 
H$_{2}$CO $3_{0,3}-2_{0,2}$, 
$3_{2,2}-2_{2,1}$,  
$3_{2,1}-2_{2,0}$),   
the sensitivities are  
$\sim$9.5 mJy\,beam$^{-1}$ per 0.17~\kms\ for 
the first six lines 
and $\sim$3.5 mJy\,beam$^{-1}$ per 1.3~\kms\ for 
the last six lines (see Table~\ref{tab:lines}). 
The 12m and 7m array line emission was combined 
with the TP observations through the feathering technique. 
All images shown in this paper are prior to the primary 
beam correction, while all measured fluxes are corrected 
for  the primary beam attenuation.

\section{Results and Analysis} 
\label{sec:results}
At the moment, the ALMA TP antennas do not provide 
continuum emission observations. Therefore, our 
analysis of continuum  and molecular lines is mostly 
focused on combined 12 m and 7 m images 
(hereafter 12m7m), 
whereas  the combined 12 m, 7 m, and TP images 
(hereafter 12m7mTP) data are also used to assess the 
missing flux in images without the total power data.

The astropy \texttt{astrodendro} package 
\citep{Rosolowsky-2008,2013A&A...558A..33A} was 
employed to identify the embedded dense cores for 
each clump (Paper~\1).  
A minimum value of 2.5$\sigma$, step of 1.0$\sigma$, 
and a minimum number of pixels equal to those 
contained in half of each synthesized beam were used 
to extract the dense cores (i.e., the leaves in the 
terminology of \texttt{astrodendro}). To eliminate 
spurious detections, we only considered the cores with 
integrated flux densities $>$3.5$\sigma$
(see Section 4.2 in Paper~\1 for more details on 
identification of dense cores).

Figure~\ref{fig:mom0}  shows the velocity-integrated 
intensity (also known as 0th-moment) maps of 
\n2dp, \dcop, DCN, \h2co ($3_{0,3}-2_{0,2}$), \ch3oh, 
and \c18o for G014.492--00.139 (hereafter G14.49), 
but excluding \ccd, $^{13}$CS,  CO, and SiO. 
The 0th-moment images of molecular lines for 
the remaining clumps are presented in Appendix~\ref{app:fig}. 
Among the three \h2co transitions,  we focus on the 
\h2co ($3_{0,3}-2_{0,2}$) line, unless otherwise noted. 
\h2co ($3_{0,3}-2_{0,2}$) traces cold dense gas better 
than the other two transitions, $3_{2,2}-2_{2,1}$ and  
$3_{2,1}-2_{2,0}$, which preferentially trace warm 
dense gas.      
The \ccd\  and $^{13}$CS lines are only detected in very 
limited regions toward 10 dense cores.  
The spatial distributions of the CO and SiO lines can be 
found in paper \2.  Both the CO and C$^{18}$O lines 
show significant emission throughout the clumps, 
albeit there is a significant depletion in their 
emission toward some dense cores. 
Both \n2dp and \dcop lines are preferentially present 
around or toward the dense cores. 
\ch3oh and \h2co emission lines are frequently found 
around outflows and dense cores. 
The majority of DCN line emission appears toward the 
protostellar cores.

\subsection{Detection Rates} 
\label{sec:detection}
In general, the emission from deuterated species is 
weak. For all the detected lines of interest, the 
spectra are averaged inside the dendrogram leaf 
(see Paper \1) that defines each core in order to 
increase the signal-to-noise ratio (S/N). We derive 
the line central velocity ($v_{\rm LSR}$), observed 
velocity dispersion 
($\sigma_{\rm obs}$ = FWHM/$2\sqrt{2\rm ln2}$), 
and peak intensity ($I$) from Gaussian fittings to 
the core-averaged spectrum for each line, 
except for \n2dp that is fitted with hyperfine  
structure (hfs) model. The derived $\sigma_{\rm obs}$ 
is not corrected by the smearing effect due to the 
channel width. 
To increase the S/N of weak line emission, the 
core-averaged spectrum is spectrally smoothed over 
2 native channels prior to Gaussian/hfs fittings if it 
shows marginal $\sim$3$\sigma$ confidence in the 
native spectral resolution. The best-fit parameters,  
that are used to compute column densities of 
molecules (see Appendix~\ref{app:column}) and 
the following analyses, are summarized in  
Appendix~\ref{app:fig}.

The CO emission is detected in all the identified cores. 
\c18o is detected in 267 out of the 294 cores, with a 
detection rate of 89\%. \h2co is the third most 
commonly detected line, which is detected in 156  
out of the 294 cores, with a detection rate of 52\%.  
\ch3oh is detected in 51 out of the 294 cores, 
leading to a detection rate of 17\%. 
Among the detected deuterated species, 
\dcop is the most commonly detected line with a 
detection rate of 39\% (116/294). \n2dp is detected 
in 54 out of the 294 cores, resulting in a detection 
rate  of 18\%. There are 7 cores associated with the 
DCN line emission, with a detection rate of  2\%. 
There is weak \ccd\ emission in 3 dense cores, with a 
detection rate of 1\%.  
$^{13}$CS is detected toward 4 dense cores, 
with a detection rate of 1\%.  
Based on the core-averaged spectra, the SiO emission 
is  detected in 27 cores.

\begin{figure*}[!ht]
\centering
\includegraphics[scale=0.235]{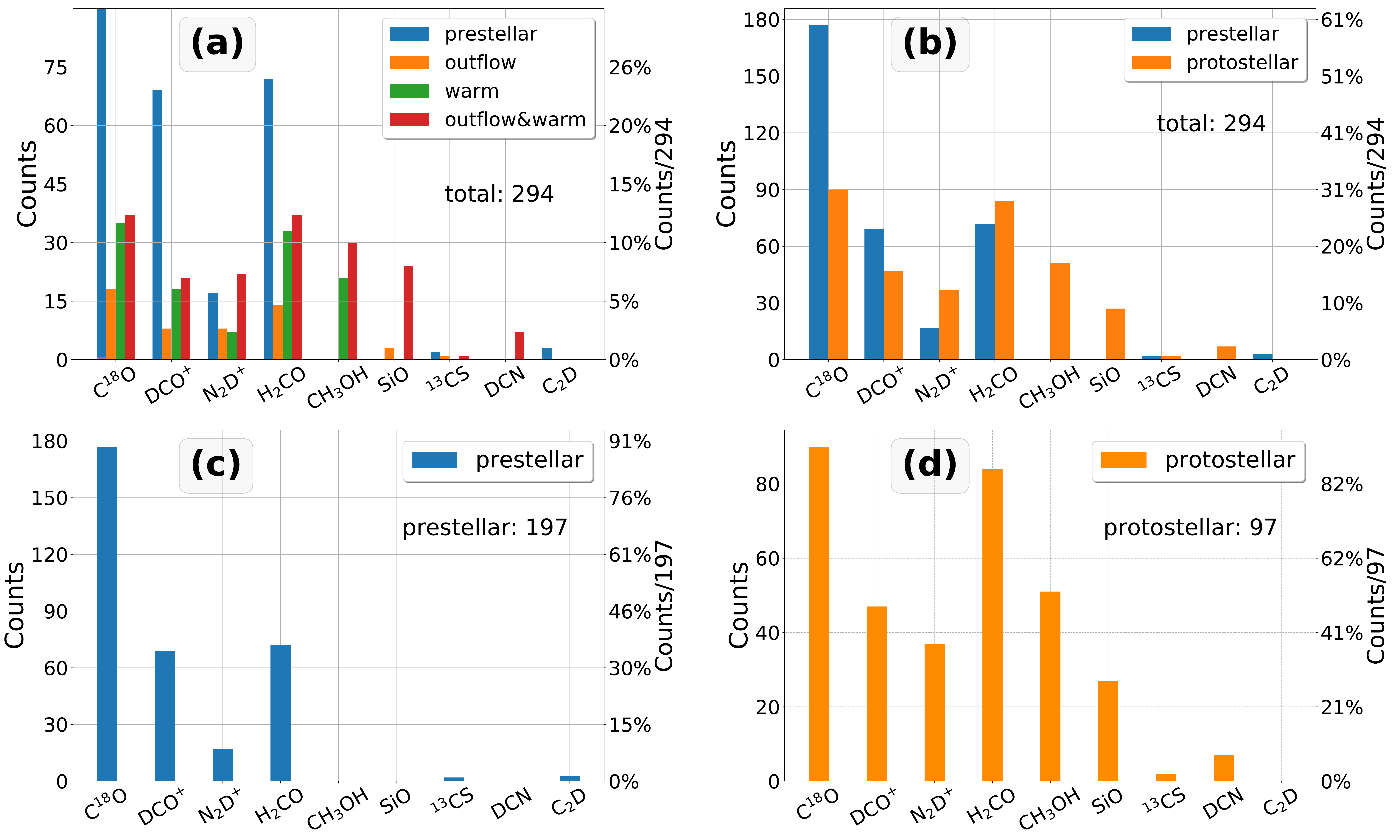}
\caption{
Histograms of number and detection rate distributions 
of detected lines for: each evolutionary category 
of cores (Panel a);  prestellar vs protostellar cores 
(Panel b); the prestellar cores only (Panel c); 
the protostellar cores only (Panel d).
}
 \label{fig:hist}
\end{figure*}

Figure~\ref{fig:hist} shows the histograms of the number 
distributions and detection rates of detected lines for the 
prestellar, protostellar cores, and for each core category. 
\n2dp is detected in  17 prestellar cores and 37 protostellar 
cores, resulting in a detection rate in the prestellar phase 
lower (9\% = 17/197) than that in the protostellar phase 
(38\% = 37/97).  The high \n2dp detection rate for the 
protostellar cores indicates that N$_{2}$D$^{+}$ does not 
exclusively trace prestellar cores and that it also frequently 
appears in the protostellar cores \citep{2019A&A...621L...7G}.  
Among the detected deuterated molecules, the \dcop has 
the highest detection rate in both the prestellar and 
protostellar cores. The \dcop detection rates are 
35\% (69/197) and 49\% (47/97) for the prestellar 
and protostellar cores  (Figure~\ref{fig:hist}),
respectively.  This indicates that  \dcop is more 
frequently detected than \n2dp in both the prestellar and 
protostellar cores.

The \h2co detection rate in the protostellar cores 
(87\% = 84/97) is more than twice of that in the 
prestellar cores (37\% = 72/197).  On the other hand, 
\h2co has a higher detection rate than \n2dp in both 
the prestellar and protostellar cores, indicating that 
\h2co is more commonly seen than the \n2dp emission 
in these dense cores at early evolutionary 
phases. There are 51 (53\% = 51/97) 
protostellar cores showing \ch3oh line emission. 
DCN is detected in 7 protostellar cores. \ccd\ is 
detected in 3 prestellar cores. $^{13}$CS is detected 
in 2 prestellar and 2 protostellar cores. The detection 
numbers and detection rates of each line in all of the 
categories are summarized in Table~\ref{tab:detection}.

\begin{figure*}[!ht]
\centering
\includegraphics[scale=0.33]{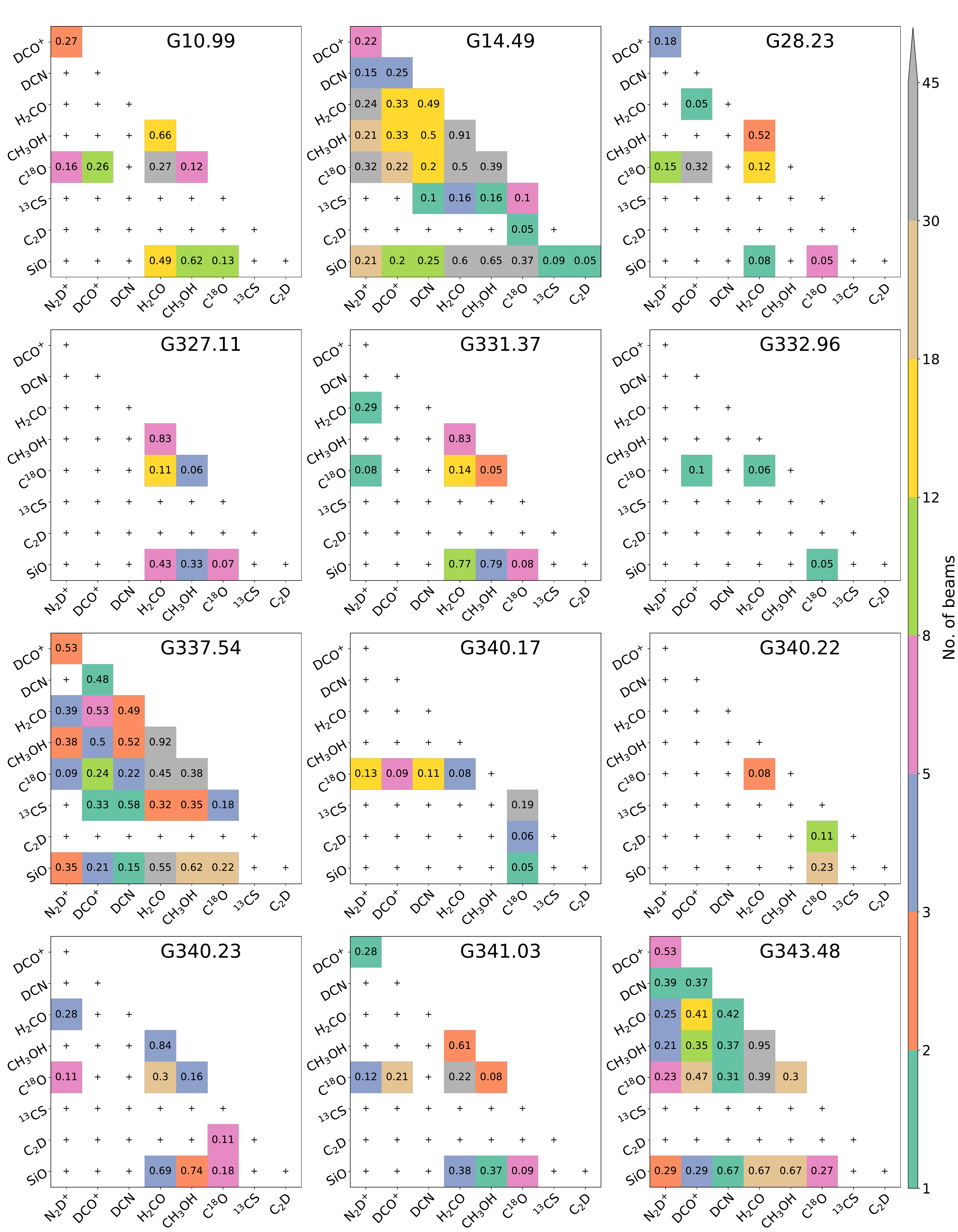}
\caption{
Integrated intensity correlation matrix, showing cross 
correlations for each pair of molecules. 
The correlation coefficients are described in the text.   
The `+' symbols mean that correlation coefficients can 
not be derived,  because of either without detection or 
the spatially overlapped emitting area is  smaller 
than 1 beam size. 
The color indicates the number of beam areas of the 
spatial overlapped emitting regions.  
A larger number of beam areas has a more robust 
derived correlation coefficient than a smaller one. 
}
 \label{fig:cross}
\end{figure*}

\subsection{Spatial Distributions of Line Emission} 
\label{sec:distrition}
In general,  both CO and \c18o show the most spatially 
extended emission over all clumps, followed by \h2co 
and \ch3oh, and then by \dcop, \n2dp, SiO, DCN, 
$^{13}$CS, and \ccd.  From Figure~\ref{fig:mom0} 
(see also Appendix~\ref{app:fig}), we note that the 
\c18o emission is extended in the IRDC clumps. 
There is very weak  \c18o emission toward some of 
the dense cores likely due to depletion 
(e.g., cores \#28/30 in G14.49; see also 
Sabatini et al. 2022, subm).

\n2dp shows extended emission toward the dense 
cores.  Peaks of the \n2dp emission are offset from the 
continuum peaks in some cores, with a significant decrease 
in intensity toward the continuum emission peaks 
(e.g., cores \#8/2 in G14.49; see Figure~\ref{fig:mom0}).  
Both N$_{2}$ depletion and CO evaporation toward the 
central of the cores can lead to decrease \n2dp abundance. 
We are unable to distinguish between these two 
possibilities with the current data. 
\dcop also shows extended emission toward the 
dense cores. However, the spatial distribution of \dcop 
does not always coincide with that of \n2dp  \citep[see, for example,][]{2022ApJ...925..144S}.  
For instance, \dcop shows a significantly different spatial  
distribution from what is  observed for \n2dp around cores 
\#7/10/21 in G14.49 (see Figure~\ref{fig:mom0}).

The peaks of the \h2co and \ch3oh emission appear to either 
coincide with the continuum peaks or locate at the direction 
of outflows in the majority of cases (Figures~\ref{fig:mom0} 
and Appendix~\ref{app:fig}). The \ch3oh emission shows 
a  behavior similar to what is observed for \h2co, but 
with less extended emission.  
This is most likely due to the fact that 
the excitation conditions of \h2co ($3_{0,3}-2_{0,2}$) 
are different from those of \ch3oh (e.g., lower upper 
level energies and critical densities; 
see Table~\ref{tab:lines}).

There are some pairs of molecules showing similar spatial 
distribution in some clumps, e.g., \h2co and \ch3oh. 
To compare the spatial trends of different species, 
we performed a 1/2-beam sampling for the 0th-moment 
of each molecular line. Each data point (or pixel) 
is about 1/2 beam width, to ensure that data points are 
not over sampled.   
The similarity of the 0th-moment maps of different 
molecules can be evaluated quantitatively by the cross 
correlation between each pair of maps according to 
\citep{2018ApJS..236...45G}
\begin{equation}
\label{eq:spcor}
\rho_{12} = \frac{\sum\limits_{i,j} I_{1,ij} \, I_{2,ij} \, w_{ij}}{\left(\sum\limits_{i,j} I_{1,ij}^{2}  \, w_{ij} \, \sum\limits_{i,j} I_{2,ij}^{2} \, w_{ij}\right)^{1/2}},
\end{equation}
where the sums are taken over all positions. 
$I_{1,ij}$ and $I_{2,ij}$ are the integrated intensities 
at the position $i$, $j$ for two arbitrary species 1 
and 2, respectively,   and the weight $w_{ij}$ is 
equal to 0 or 1 depending on whether or not  the line 
emission was detected at that position. $\rho_{12}$ is 
equal to 1 if the 0th-moment maps of two molecules 
have the same spatial distribution. Although some 
pairs of molecules that show weak emission  do not 
have statistically significant number of independent 
data points, it is still worth an examination. 
To alleviate the effect of insufficient numbers of 
independent data points for comparison,  we avoid 
deriving the correlation coefficient of the pair of 
molecules whose overlapped emitting area is smaller 
than 1 synthesized beam.

Figure~\ref{fig:cross} presents cross correlation 
coefficients for the 0th-moment map of each pair of 
molecules. The correlation between \n2dp and \dcop is 
better than those of the other detected deuterated 
molecules (i.e., DCN and \ccd) in all the clumps. 
This is because the emitting regions of DCN and 
\ccd\ lines are smaller than either \n2dp or \dcop 
(Figure \ref{fig:mom0}).  The \dcop line emission 
coincides better with \h2co and \ch3oh than \n2dp 
in terms of spatial distribution.   
This may be because \dcop, \h2co and \ch3oh
have a common precursor molecule of CO  that 
tends to destroy \n2dp (see Sections~\ref{sec:deu} 
and \ref{sec:h2co}).

In general, the \ch3oh and \h2co emission have the most 
similar spatial distributions in integrated intensities 
with the highest correlation coefficient among the 
detected lines.  Most of the \ch3oh and \h2co emission 
appear to be around either outflows or dense cores 
(Figure~\ref{fig:mom0}). In addition, both the \ch3oh 
and \h2co emission show a good spatial correlation with 
the SiO emission in most of the clumps 
(Figure~\ref{fig:cross}). These results suggest 
that both the \ch3oh and \h2co line  emission are 
closely related to outflow activity toward 
protostellar cores in the ASHES clumps.

\begin{figure*}[!ht]
\centering
\includegraphics[scale=0.22]{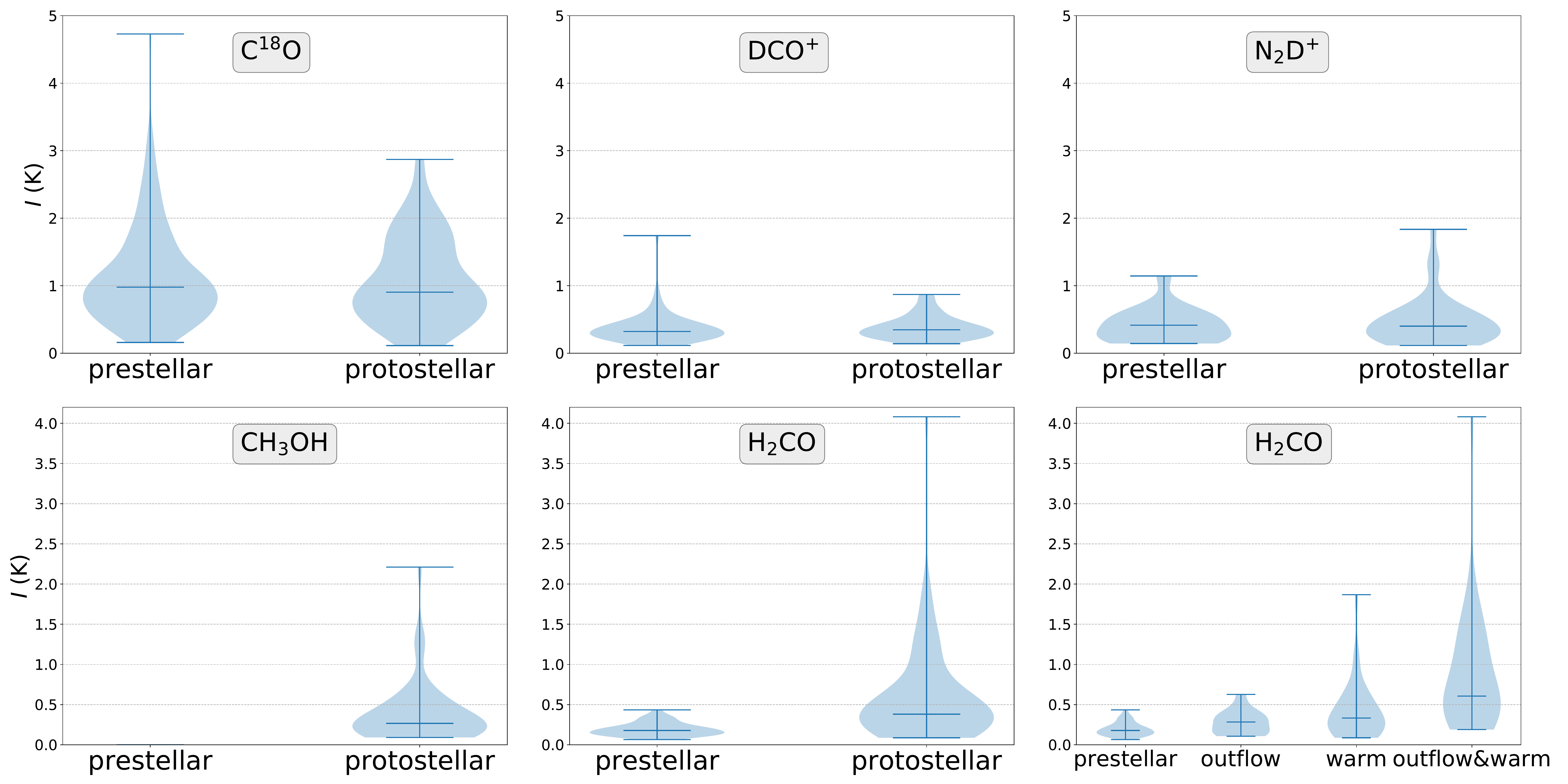}
\includegraphics[scale=0.22]{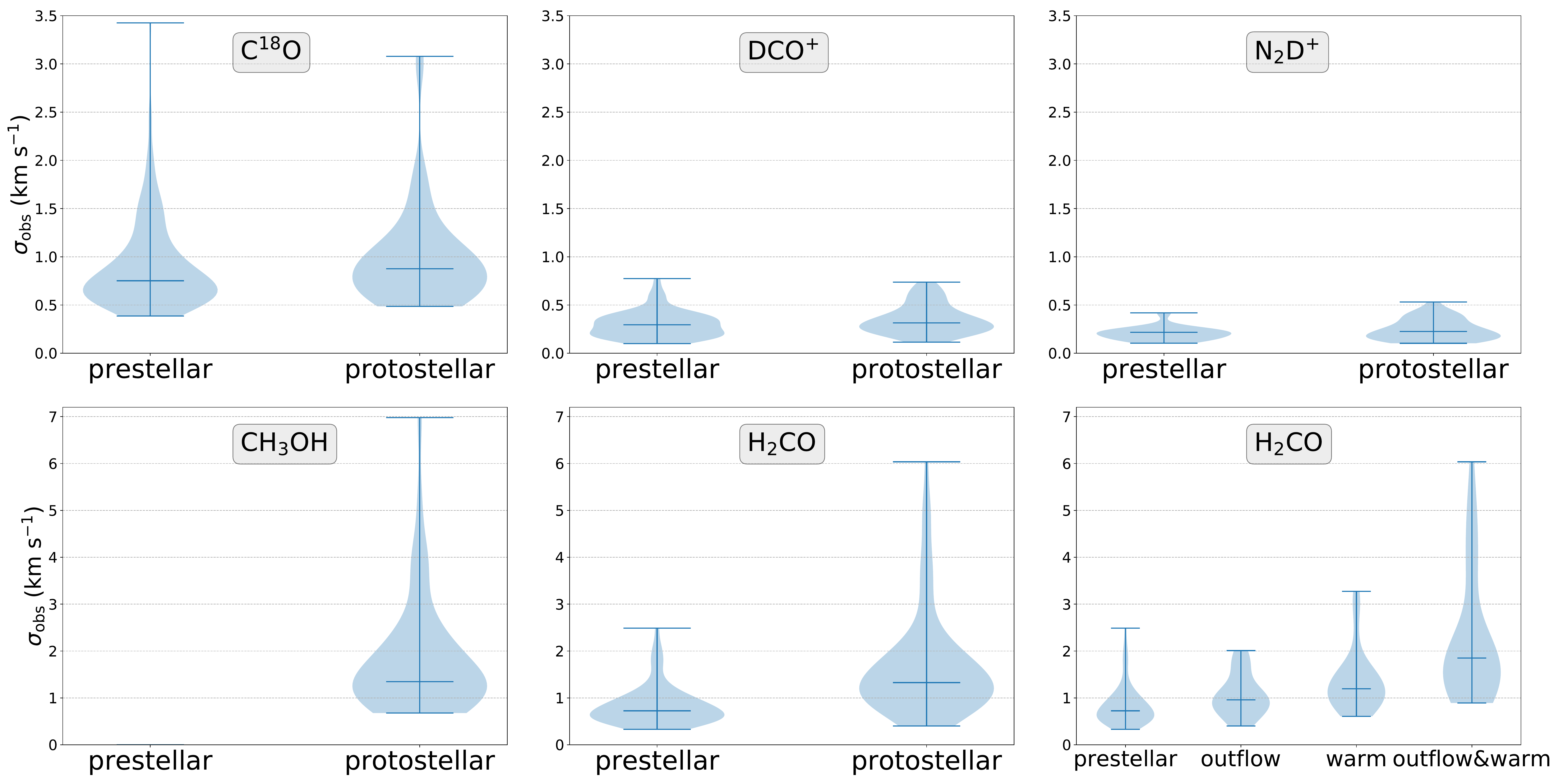}
\caption{
Violin plots of the peak intensity $I$ and observed 
velocity dispersion $\sigma_{\rm obs}$ distributions 
for each line. 
The shape of each distribution shows the probability 
density of the data smoothed by a kernel density 
estimator. 
The last panel shows the \h2co peak intensity 
distributions of each category of the dense cores. 
The blue horizontal bars from the top to bottom 
in each violin plot represent the maximum, mean, 
and minimum values, respectively. 
}
 \label{fig:Ipk}
\end{figure*}

\subsection{Molecular Line Parameters} 
\label{sec:param}
Figure~\ref{fig:Ipk} shows the distribution of the 
derived $I$ and $\sigma_{\rm obs}$ of each line for 
the prestellar and protostellar cores (see also 
Table~\ref{tab:mean}). 
Overall, $\sigma_{\rm obs}$ shows no  significant  
difference between the prestellar and protostellar 
cores for the \c18o and \dcop emission. 
This indicates that the protostellar cores are 
still at a very early evolutionary phase, in which 
the line widths of \c18o (a low density gas tracer) 
and \dcop (a high density tracer) have not been 
significantly affected by  protostellar activity.  
In general, \n2dp line  has comparable $I$ in both 
the prestellar and protostellar cores, except for 
4 protostellar cores that present a relatively higher 
$I$ than that of prestellar cores. 
On the other hand, \n2dp line emission shows 
relatively larger $\sigma_{\rm obs}$ in the 
protostellar cores compared with  the prestellar cores. 
This might be because the \n2dp line emission toward 
protostellar cores is influenced by the injection of 
turbulence as a result of protostellar activity.

For \h2co emission, both $I$ and $\sigma_{\rm obs}$ in the protostellar cores 
are higher than in the prestellar cores. This is because \h2co abundances 
can be significantly enhanced in warm and dense 
environments, and its line width can be broadened 
by protostellar activity  
\citep[e.g., outflow;][]{2010A&A...522A..91T,
2012ApJ...754...70S}.  
In addition, \h2co is the molecule in our sample that shows a 
clearly increasing trend in both $I$ and $\sigma_{\rm obs}$ 
from categories 1 to 4 (see Figure~\ref{fig:Ipk}), indicating 
that its abundance and line width are sensitive to the evolution 
of the dense cores.  This implies that \h2co could be used 
as a diagnostic tool to infer star formation activities.

From Figure~\ref{fig:Ipk}, one notes that the \h2co 
($\langle \sigma_{\rm obs} \rangle$ = 1.68 \kms) and 
\ch3oh ($\langle \sigma_{\rm obs} \rangle$ = 1.82 \kms) 
lines show relatively larger $\sigma_{\rm obs}$ 
than \c18o ($\langle \sigma_{\rm obs} \rangle$ = 0.99 \kms), 
\dcop ($\langle \sigma_{\rm obs} \rangle$ = 0.35 \kms), 
and \n2dp ($\langle \sigma_{\rm obs} \rangle$ = 0.25 \kms) 
toward the protostellar cores (see also Table~\ref{tab:mean}).  
This could be because \h2co and \ch3oh are 
associated with more turbulent gas components affected 
by protostellar activity 
\citep[e.g., outflow;][]{2021A&A...655A..65T}. 
We refrain from investigating $I$ and $\sigma_{\rm obs}$ 
for the $^{13}$CS, DCN, and \ccd emission due to a lack 
of a sufficient number of detections for a meaningful 
analysis, as well as the SiO emission that is mainly 
associated with outflows.

\begin{figure*}[!ht]
\centering
\includegraphics[scale=0.21]{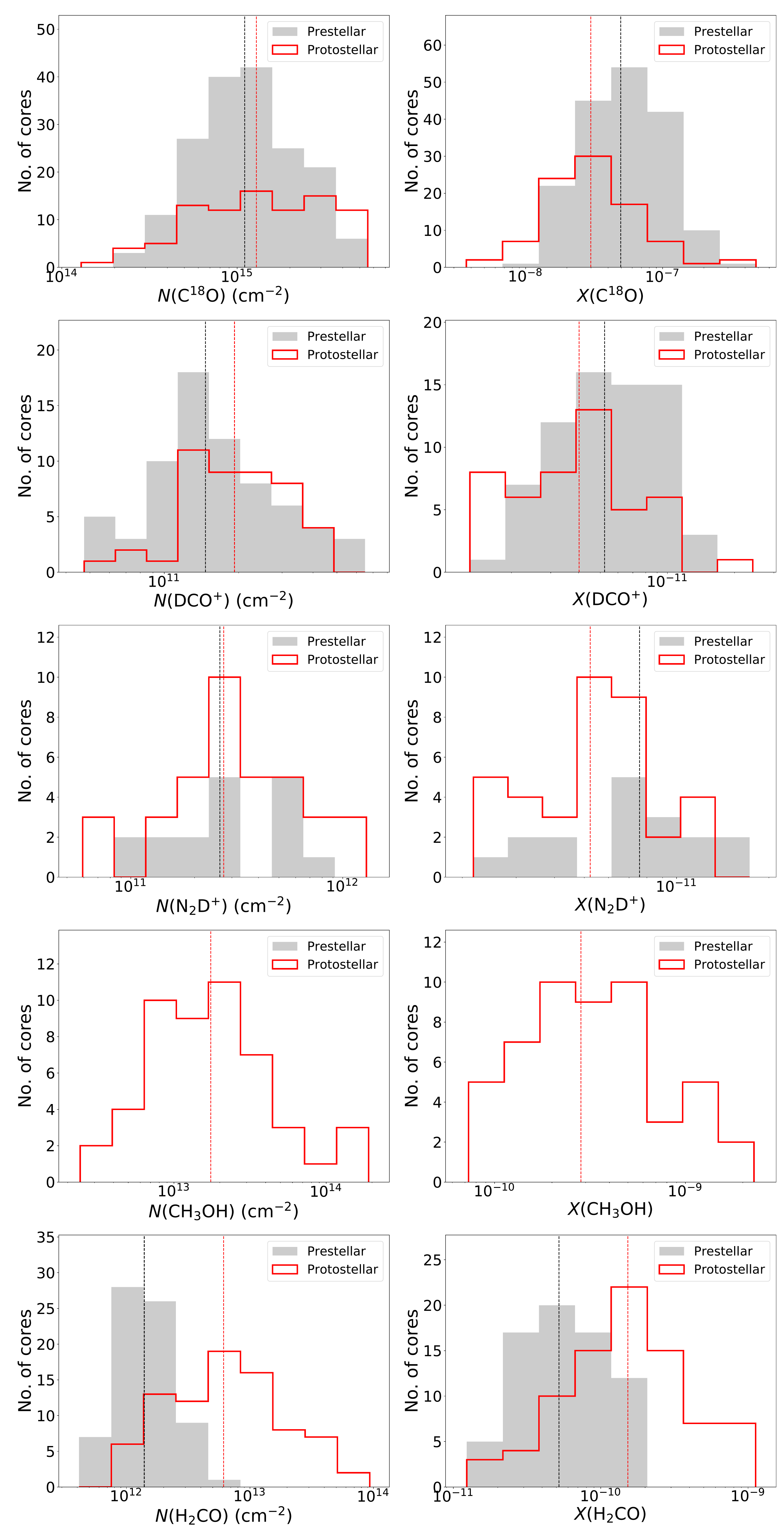}
\caption{
Left column: histograms of the molecular column 
densities for the prestellar and protostellar cores. 
Right column: histograms of the molecular abundances   
for the prestellar and protostellar cores. 
The black dashed vertical lines indicate the median 
values of the parameters.
}
 \label{fig:NX}
\end{figure*}

\subsection{Derivation of Physical Parameters} 
\label{sec:density}
The rotational excitation temperature ($T_{\rm NH_3}$) is  
derived from NH$_{3}$ (1, 1) and (2, 2) transition lines 
obtained from the CACHMC survey (Complete 
ATCA\footnote{The Australia Telescope Compact Array} 
Census of High-Mass Clumps; Allingham et al., 
2022 in prep)  at $\sim$5\arcsec angular resolution 
(see Appendix~\ref{app:tnh3} for detailed procedure 
on excitation temperature determination). 
More details procedure on temperature determination 
and survey results will be presented in a forthcoming 
paper (Allingham et al., in prep).  
The derived $T_{\rm NH_3}$ is used as excitation 
temperature in the calculation of all molecular parameters,  
except for G332.96 that has no available NH$_{3}$ data. 
We used the dust temperature at the clump scale of 
12.6~K for G332.96 (Paper \1). The averaged temperature 
of dense cores in the same category for a given  clump 
is used for those dense cores that have no available 
$T_{\rm NH_3}$.  
The approximation of using $T_{\rm NH_3}$ as the 
excitation temperature is based on the LTE conditions 
in calculation of molecular density. 
The results could be highly sensitivity to the choice of 
the excitation temperature.

Using these new temperatures, we have recalculated 
the ASHES cores properties. The analysis on how 
these new core properties affect the analysis presented in \cite{2019ApJ...886..102S} will be presented in a 
forthcoming paper (Li et al. 2022, in prep). 
For the chemical analysis of the current work, 
we have used the updated core-average 
H$_2$ column densities and core volume densities 
(see Appendix~\ref{app:column}).
The updated parameters are  
4.9\,$\times$\,10$^{21}$--2.0\,$\times$\,10$^{23}$ cm$^{-2}$ 
for the core-averaged column density and 
1.4\,$\times$\,10$^{5}$--1.7\,$\times$\,10$^{7}$ cm$^{-3}$ 
for the core-averaged volume density. 
The protostellar cores have higher column densities and 
volume densities than the prestellar cores 
(Table~\ref{tab:Nline}).

Assuming local thermodynamic equilibrium (LTE) and 
optically thin molecular emission, 
the molecular column density  
($N_{\rm mol}$) can be estimated from the velocity 
integrated intensity (see Appendix~\ref{app:column}, 
for a detailed derivation of the column density).   
To study the properties of the molecules,  
we also calculated the molecular abundances, 
$X_{\rm mol} = N_{\rm mol}/N_{\rm H_2}$, 
using the updated H$_{2}$ column density and derived 
molecular column densities.

The derived \n2dp column densities range from 
5.9\,$\times$\,10$^{10}$ to 1.3\,$\times$\,10$^{12}$ 
cm$^{-2}$,  resulting in \n2dp abundances of  
2.2\,$\times$\,10$^{-12}$--1.7\,$\times$\,10$^{-11}$.  
The \n2dp column densities are similar to the value of 
6.2\,$\times$\,10$^{11}$ cm$^{-2}$ obtained for other 
IRDCs \citep[e.g.,][]{2015A&A...579A..80G,
2016MNRAS.458.1990B,2010ApJ...713L..50C}, 
and the \n2dp 
abundances are comparable to the values of 
$\sim$10$^{-12}$ in other massive clumps 
\citep{2019A&A...621L...7G}. 
The estimated \dcop column densities  are between  
4.7\,$\times$\,10$^{10}$ and 6.5\,$\times$\,10$^{11}$ 
cm$^{-2}$, leading to \dcop abundances  of 
1.3\,$\times$\,10$^{-12}$--2.4\,$\times$\,10$^{-11}$. 
$N_{\rm DCO^{+}}$ is similar to the values reported  
in other IRDCs 
\citep[$\leqslant$3\,$\times$\,10$^{11}$ 
cm$^{-2}$][]{2015A&A...579A..80G}. 
The derived DCN column densities and abundances are   
2.7\,$\times$\,10$^{11}$--2.0\,$\times$\,10$^{12}$ cm$^{-2}$ 
and 5.6\,$\times$\,10$^{-12}$--1.2\,$\times$\,10$^{-11}$, 
respectively.  The DCN abundances are significantly lower 
than those in more evolved dense cores in W3 
\citep[$>$10$^{-10}$;][]{2020A&A...636A.118M}. 
\ccd\ has column densities of 
4.8\,$\times$\,10$^{12}$--1.1\,$\times$\,10$^{13}$ cm$^{-2}$ 
and abundances of 
1.5\,$\times$\,10$^{-10}$--4.2\,$\times$\,10$^{-10}$.

Given that these dense cores are still in very early 
evolutionary phases and are characterized by a cold 
environment, \h2co is expected to form in ice on 
the grain surface and subsequently to be released to 
the gas phase \citep{2005A&A...437..501J}. 
We assumed an ortho-to-para ratio of 1.6 which is 
consistent with thermalization at a low temperature 
of T $\sim$ 15 K \citep{2005A&A...437..501J}. 
The derived \h2co column densities and abundances 
are 4.1\,$\times$\,10$^{11}$--9.3\,$\times$\,10$^{13}$ 
cm$^{-2}$ 
and 1.2\,$\times$\,10$^{-11}$--1.1\,$\times$\,10$^{-9}$, 
respectively. \ch3oh has column densities of  
2.4\,$\times$\,10$^{12}$--1.9\,$\times$\,10$^{14}$ 
cm$^{-2}$ and abundances of 
7.3\,$\times$\,10$^{-11}$--2.3\,$\times$\,10$^{-9}$ 
for the detected dense cores. 
The abundances of  \h2co and \ch3oh are similar to 
those in dense cores prior to the hot core phase in 
other IRDCs \citep[a few 10$^{-10}$ for \h2co and 
\ch3oh;][]{2014A&A...563A..97G,2020A&A...636A.118M}, 
but lower than those in hot core and UCHII regions   
\citep[$\geqslant$10$^{-9}$; 
e.g.,][]{2014A&A...563A..97G,2020A&A...636A.118M}.

\c18o optical correction factor, 
$C_{\tau} = \tau_{\rm C^{18}O}/[1 - \rm{exp}(-\tau_{\rm C^{18}O})]$,  
is applied in the calculation of \c18o column density. 
$C_{\tau}$ is derived following the approach described 
in \cite{2019MNRAS.490.4489S}. 
The detailed estimation of \c18o correction factor 
is presented in \cite{2022arXiv220712431S}. 
The calculated \c18o column densities vary from 
1.3\,$\times$\,10$^{14}$ to 5.6\,$\times$\,10$^{15}$ 
cm$^{-2}$,  resulting in \c18o abundances  
of  3.7\,$\times$\,10$^{-9}$--4.8\,$\times$\,10$^{-7}$. 
The estimated SiO column densities vary from 
2.3\,$\times$\,10$^{11}$--3.2\,$\times$\,10$^{13}$ 
cm$^{-2}$ that are comparable to the values in outflows, 
whereas the derived abundances of 
3.0\,$\times$\,10$^{-12}$--3.4\,$\times$\,10$^{-10}$ 
are lower than those found in outflows 
($\geqslant$1.1\,$\times$\,10$^{-9}$; 
see Paper~\2).

Figure~\ref{fig:NX} shows histograms of molecular column 
densities and abundance ratio distributions for the 
prestellar and protostellar cores (see also 
Table~\ref{tab:Nline}). 
Except for \c18o and \n2dp that have a similar column 
density in the prestellar and protostellar cores, all 
other molecules have  column densities in the prestellar 
cores relatively lower than those from protostellar 
cores.    
On the other hand, \n2dp, \dcop, and 
\c18o abundances are higher in the prestellar cores 
than in the protostellar cores, indicating that the 
abundances of these molecules tend to decrease with 
the evolution of the dense cores. 
These abundance variations are dominated by the effect 
of the H$_{2}$ density increase rapidly from prestellar 
to protostellar phases (see Section~\ref{sec:deu}). 
In contrast, 
the abundance of \h2co in the protostellar cores is  
higher than in the prestellar ones, suggesting that its 
abundance can be enhanced with core evolution.

\begin{figure*}[!ht]
\centering
\includegraphics[scale=0.19]{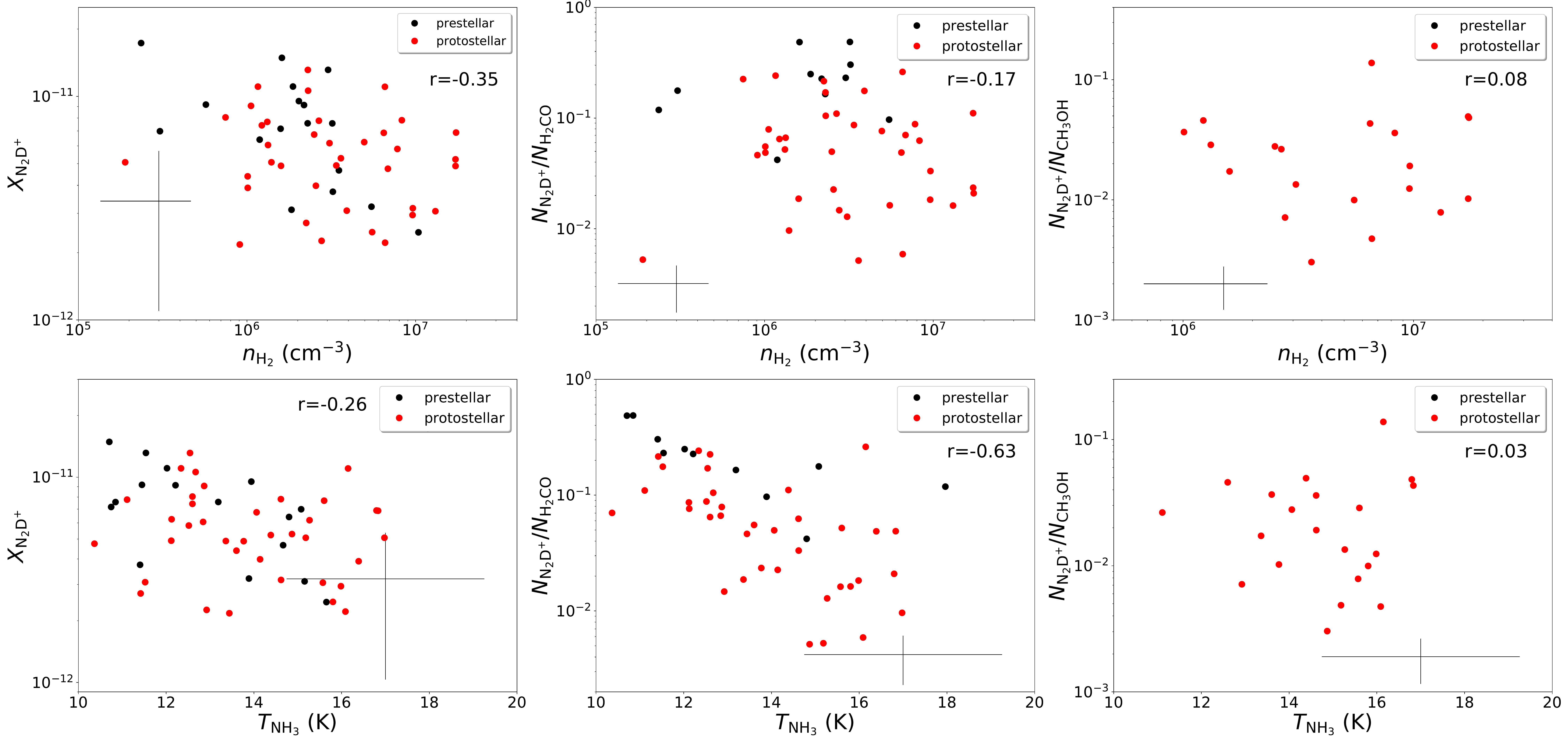}
\caption{
Top row: plots of $N_{\rm H_2}$ versus $N_{\rm N_2D^+}$, 
$N_{\rm N_{2}D^{+}}/N_{\rm H_{2}CO}$, and 
$N_{\rm N_{2}D^{+}}/N_{\rm CH_{3}OH}$. 
Bottom row: plots of  $T_{\rm NH_3}$ versus 
$X_{\rm N_2D^+}$, $N_{\rm N_{2}D^{+}}/N_{\rm H_{2}CO}$, 
and $N_{\rm N_{2}D^{+}}/N_{\rm CH_{3}OH}$. 
The correlation coefficients derived from the 
Spearman rank correlation test are texted in 
each panel. The black cross shows the typical 
uncertainty in each panel. 
}
 \label{fig:n2dp}
\end{figure*}

\begin{figure*}[!ht]
\centering
\includegraphics[scale=0.19]{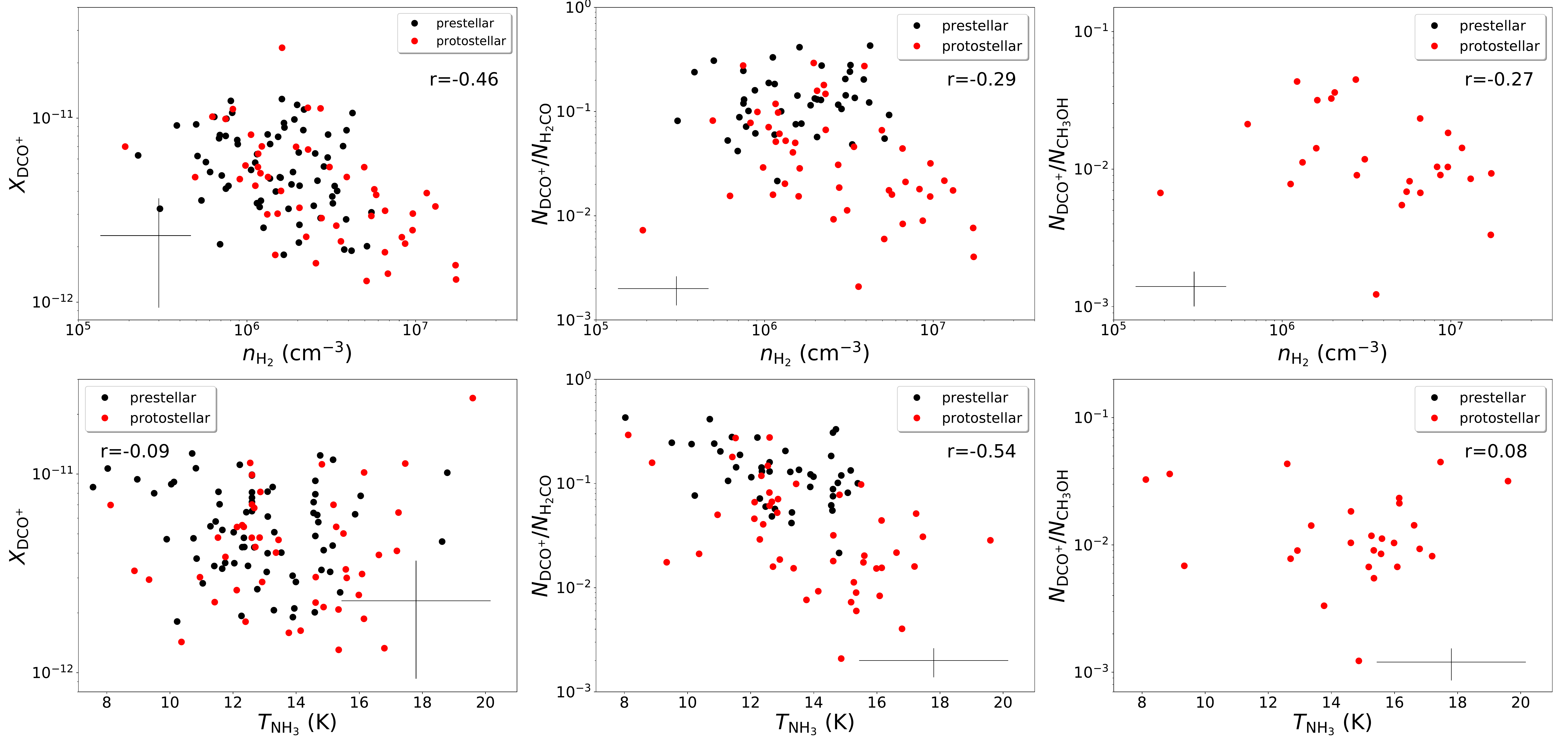}
\caption{
Same as Figure~\ref{fig:n2dp}, but for the \dcop. 
}
 \label{fig:dcop}
\end{figure*}

\section{Discussion} 
\label{sec:discu}
\subsection{Missing flux} 
\label{sec:conf}
To understand the impact of the missing flux on the 
line emission, we compared the 12m7m and 12m7mTP 
datasets. We investigated the integrated intensities, 
peak intensities, line widths, and central velocities 
of line emission between the two datasets.

For the \n2dp line emission, we find that the 12m7m 
data have recovered about 92\% of the 12m7mTP flux. 
The differences of the measured $I$  and 
$\sigma_{\rm obs}$ are a factor of 0.94 and 
0.99, respectively. 
For the \dcop line emission, the mean ratios of 12m7m 
to 12m7mTP are 0.89 for $I$  and  0.9 for 
$\sigma_{\rm obs}$, resulting a mean flux ratio 
(velocity integrated intensity) of 0.79.  The \dcop 
emission behaviour is very similar to that of \n2dp. 
These results indicate that most of the \n2dp and 
\dcop emission is compact and recovered without 
the addition of total power data from single-dish 
telescopes.

For the \h2co ($3_{0,3}-2_{0,2}$) line emission, the 
12m7m   data recover about 71\% of the 12m7mTP flux, 
and the mean ratios of 12m7m to 12m7mTP for $I$ and 
$\sigma_{\rm obs}$ are 0.82 and 0.87, respectively. 
On other hand, \ch3oh has similar flux, $I$, and 
$\sigma_{\rm obs}$ between the 12m7m and 12m7mTP 
datasets; the mean ratio (12m7m/12m7mTP)  is 0.94 for 
flux, 0.94 for $I$, and 1.0 for $\sigma_{\rm obs}$. 
This indicates that \ch3oh traces more spatially compact 
emission compared with \h2co. This suggests that the 
observed \ch3oh transition may  preferentially be 
concentrated near the protostars or in knots in outflows.

For \c18o, the mean ratios of 12m7m to 12m7mTP data are 
0.54 for the intensity peak  and  0.7 for the velocity 
dispersion, resulting a mean flux ratio of 0.36. 
Among the detected lines 
(except for CO), \c18o suffers most severely from  
missing flux.  This indicates that \c18o probes a 
significant amount of diffuse molecular gas. 
For the remaining lines with relatively low detection 
rates,  their 12m7m images are weakly affected by 
missing flux. The mean flux ratios of 12m7m to 
12m7mTP are 1.0 for SiO, 0.96 for $^{13}$CS, 0.9 
for DCN, and 0.98 for \ccd.

\begin{figure}[!ht]
\centering
\includegraphics[scale=0.3]{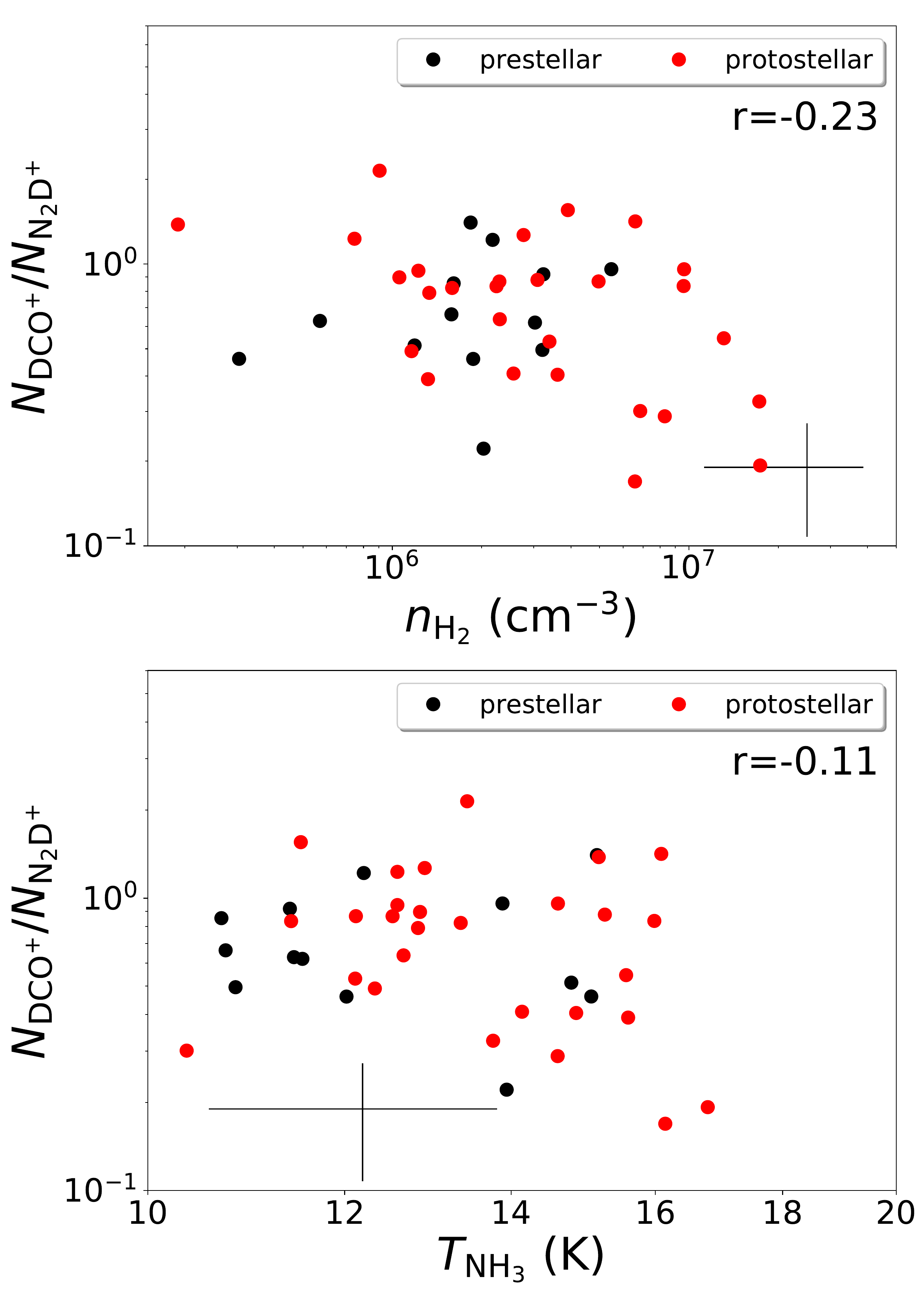}
\caption{
The column density ratio of \dcop to \n2dp 
versus the core-averaged volume density $n_{\rm H_2}$ 
and the temperature $T_{\rm NH_3}$.  
}
 \label{fig:dcopn2dp}
\end{figure}

\subsection{Deuterated Molecules} 
\label{sec:deu}
\n2dp is expected to be abundant in cold ($<$20 K) 
and dense ($\sim$10$^{5} \rm \, cm^{-3}$) regions, in which 
its major destroyer, CO, is significantly depleted onto grain 
surfaces. \n2dp can be formed via N$_{2}$ reacting with 
H$_{2}$D$^{+}$ (dominant reaction:  
N$_{2}$ + H$_{2}$D$^{+}$ $\rightarrow$ \n2dp + H$_{2}$),  
D$_{2}$H$^{+}$, or D$_{3}$$^{+}$ \citep{2009A&A...494..623P}, 
and destroyed by CO 
(\n2dp + CO $\rightarrow$ \dcop + N$_{2}$) or electrons  
\citep[\n2dp + e$^{-}$ $\rightarrow$  ND +N or 
D + N$_{2}$;][]{2022ApJ...925..144S}.     
\dcop is also considered as a cold and dense gas tracer. 
However, \dcop requires gas-phase CO for its formation 
at cold temperatures (T $<$ 20--50 K),  i.e., 
H$_{2}$D$^{+}$ + CO $\rightarrow$ \dcop + H$_{2}$ 
\citep{1987ApJ...317..220W}.   
Therefore, the \dcop abundance does not decrease rapidly 
when CO is released to the gas phase from the grain 
mantles \citep{1984ApJ...287L..47D}.  
The different chemistry between \n2dp and \dcop may 
explain their different spatial distributions seen in 
Figure~\ref{fig:mom0} \citep[e.g.,][]{2022ApJ...925..144S}. DCN and \ccd tend to form in 
warm environments \citep{2001ApJS..136..579T,
2013ApJS..207...27A}.

\n2dp and \dcop column densities exhibit a strong 
correlation (see Figure~\ref{fig:Nmol} in 
Appendix~\ref{app:fig}), with a Spearman's rank correlation 
coefficient\footnote{Spearman rank correlation test 
is a nonparametric measure of the monotonicity of 
the relationship between two variables. 
The correlation coefficient $|r|\geqslant 0.5$ means 
strong correlation, $0.5 > |r|\geqslant 0.3$ means 
moderate correlation, $0.3 > |r|\geqslant 0.1$ means 
weak correlation, and $0.1 > |r|$ means no correlation 
\citep{Cohen1988}.} 
of $r$ = 0.58. This is mostly because both lines 
trace cold and dense gas in the dense cores. 
The \n2dp abundance appears to decrease with increasing 
volume densities $n_{\rm H_2}$, with a moderate correlation 
coefficient of -0.35 (see Figures~\ref{fig:n2dp}). 
As mentioned in Section~\ref{sec:density}, the volume 
density increases with the evolution of the dense cores 
that continue to accumulate mass via accretion of material 
from the natal clumps. 
The $n_{\rm H_2}$--$X_{\rm N_2D^+}$ anticorrelation 
suggests that the \n2dp abundance decreases with the 
core evolution.   
As the cores evolve, the gas temperature increases, thereby 
lowering the deuterium enhancement. In this case, one 
naively expects an anticorrelation between \n2dp 
abundance and $T_{\rm NH_3}$.  As seen in 
Figures~\ref{fig:n2dp}, there is no obvious trend 
between the \n2dp abundance and $T_{\rm NH_3}$. 
These cores are still very cold with temperatures between 
10 and 20 K,  which are lower than the sublimation temperature 
of CO ($\sim$25 K). This may explain why the \n2dp 
abundance does not vary significantly with $T_{\rm NH_3}$. 
In addition, this suggests that these cores are still at 
a very early evolutionary phase, in which the surrounding  
environment has  not been significantly warmed up by the 
young stellar objects (YSOs).  
On the other hand, we cannot completely rule out the 
possibility that $T_{\rm NH_3}$ does not accurately reflect 
the true temperature of the dense molecular gas traced 
by \n2dp, because the spatial resolution of $T_{\rm NH_3}$ 
($\sim$5\arcsec) is coarser than the ALMA observations 
($\sim$1.2\arcsec) and the critical density of NH$_3$ (1,1) 
is only of a few 10$^4$ cm$^{-3}$. Determination of 
temperatures at $\sim$1\arcsec scales would be necessary 
to verify whether the $T_{\rm NH_3}$ varies significantly 
down to the small scale. 
The $n_{\rm H_2}$--$X_{\rm N_2D^+}$ and  
$X_{\rm N_2D^+}$-- $T_{\rm NH_3}$ results suggest that the 
\n2dp abundance variation is dominated by the H$_{2}$ core 
density in the ASHES sample.

The column density ratio of $N_{\rm N_2D^+}$/$N_{\rm H_2CO}$ 
appears to decrease as a function of increasing $n_{\rm H_2}$.  
However, a closer look reveals that this weak anticorrelation  
is mostly dominated by the protostellar cores.  Both 
$N_{\rm N_2D^+}$ and $N_{\rm H_2CO}$ tend to increase with 
$n_{\rm H_2}$, and therefore, the 
$N_{\rm N_2D^+}$/$N_{\rm H_2CO}$-$n_{\rm H_2}$ 
anticorrelation implies that $N_{\rm H_2CO}$ is  more 
sensitive to $n_{\rm H_2}$ than $N_{\rm N_2D^+}$ toward 
the protostellar cores (see also Section~\ref{sec:param}). 
In addition, the ratio $N_{\rm N_2D^+}$/$N_{\rm H_2CO}$ 
also appears to decrease with increasing $T_{\rm NH_3}$, 
with a correlation coefficient of -0.63. This is because the 
\h2co abundance is enhanced by elevated temperatures. 
The $N_{\rm N_2D^+}$/$N_{\rm CH_3OH}$ ratio shows 
no clear trend with $n_{\rm H_2}$ or $T_{\rm NH_3}$. 
In general, \dcop shows a similar trend to those seen in 
\n2dp,  except for 
$n_{\rm H_2}$--$N_{\rm DCO^+}$/$N_{\rm CH_3OH}$ 
($r$ = -0.27) that presents a weak anticorrelation.   
The difference in \n2dp and \dcop may be a result  
of different chemistry.

There is a weak anticorrelation between $n_{\rm H_2}$ 
and the $N_{\rm DCO^+}$/$N_{\rm  N_2D^+}$ ratio 
(see Figure~\ref{fig:dcopn2dp}), with a 
correlation coefficient of $r$ = -0.23. 
From Figures~\ref{fig:n2dp} and \ref{fig:dcop}, one 
notes that both \n2dp and \dcop abundances decrease 
with increasing $n_{\rm H_2}$, thereby causing the 
ratio of \dcop to \n2dp to slightly change with 
$n_{\rm H_2}$. 
In addition, the $N_{\rm DCO^+}$/$N_{\rm  N_2D^+}$ 
ratio shows only minor changes with $T_{\rm NH_3}$, 
which gives a correlation coefficient of $r$ = -0.11. 
This can be ascribed to the fact that the both \dcop 
and \n2dp abundances do not vary significantly with 
$T_{\rm NH_3}$ (see Figures~\ref{fig:n2dp} and 
\ref{fig:dcop}).

\subsection{H$_{2}$CO and CH$_{3}$OH} 
\label{sec:h2co}
\ch3oh is considered to be mostly formed on the surface 
of dust grains via the hydrogenation of CO with some 
intermediate products 
\citep[e.g., \h2co;][]{1997ApJ...482L.203C,
2004ApJ...616..638W,2009A&A...505..629F}, 
\begin{equation}
\label{eq:ch3oh}
\rm CO \; \xrightarrow[\text{H}]  \; 
HCO \xrightarrow[\text{H}] \; 
H_{2}CO \xrightarrow[\text{H}] \; 
CH_{2}OH, CH_{3}O \xrightarrow[\text{H}]\;
CH_{3}OH.
\end{equation}  
Unlike \ch3oh, which formed entirely on the surfaces 
of dust grains, \h2co can be formed efficiently through 
both grain-surface reaction in cold environments and 
gas-phase reactions in warm/hot environments 
\citep{2000A&AS..146..157L,2006FaDi..133...51G,
2009A&A...505..629F}.

\ch3oh shows a dip in its emission distribution at the 
center of some  dense cores, such as the protostellar 
core \#15 in G337.54 (Appendix~\ref{app:fig}). 
A similar feature has been observed in low-mass prestellar  
cores, e.g., L1498 and L1517B \citep{2006A&A...455..577T}. 
These authors suggest that the weaker \ch3oh emission 
toward the center of the dense cores is due to the 
significant \ch3oh depletion in the cold and dense 
environments. Therefore, the low detection rate of 
\ch3oh in the dense cores may be partly ascribed 
to \ch3oh depletion.

Among the detected molecules, \h2co and \ch3oh column 
densities show the strongest correlation, with a 
correlation coefficient of 0.83  
(see Figure~\ref{fig:Nmol} in Appendix~\ref{app:fig}). 
As mentioned in Section~\ref{sec:param}, both \h2co 
and \ch3oh tend to probe the 
dense and high velocity molecular gas, and hence, the 
protostellar activity may be partly responsible for the 
strong correlation of the two species. For example, 
outflow/shocks can release both \h2co and \ch3oh from 
grain mantles to the gas phase.  
In addition, both \h2co and \ch3oh column densities 
show strong correlation with the SiO column density 
(see Figure~\ref{fig:Nmol} in Appendix~\ref{app:fig}); 
the correlation coefficient is 0.73 for 
$N_{\rm H_{2}CO}$--$N_{\rm SiO}$ and 0.84 for 
$N_{\rm CH_{3}OH}$--$N_{\rm SiO}$.

The derived abundance ratios of \h2co/\ch3oh = 0.1--2.3  
are comparable to the values of 0.9--2.5 reported in 
dense clumps 
\citep[$n_{\rm H_2} \sim$ 10$^{6}$ 
cm$^{-3}$;][]{2010A&A...511A..82L}, hot-corinos  
\citep[0.7--4.3;][]{2004A&A...416..577M,
2005A&A...442..527M}, and low-mass starless cores 
\citep[1.1--2.2; e.g., L1498 and 
L1517B;][]{2006A&A...455..577T}. 
These values are higher than the abundance ratios  
derived in hot cores \citep[0.13--0.28;
][]{2007A&A...465..913B} 
and shocked gas in the Galactic Center clouds 
\citep[0.01-0.1;][]{2006A&A...455..971R,
2021ApJ...909..177L},  
whereas they are significantly lower than those in 
the inter-clump medium 
\citep[\h2co/\ch3oh =  14--1400; 
$n_{\rm H_2} \sim$ 10$^{4}$ 
cm$^{-3}$;][]{2010A&A...511A..82L}. 
The discrepancies could be attributed to either 
different dominant formation mechanisms or 
different chemical conditions. A detailed 
chemical modeling and observational comparison 
is needed to distinguish between these 
possibilities, which is beyond the scope 
of this paper.

\section{Conclusion}
\label{sec:conclu}
In this paper, we analyze ALMA data from the 
ASHES project to investigate the chemistry of  
294 dense cores in 12 massive 70 \um dark clumps. 
We have studied the spatial distributions and 
chemical variations of \c18o, \dcop, \n2dp, 
DCN, \ccd\, \h2co, \ch3oh, $^{13}$CS, and SiO 
in different evolutionary phases of dense cores.
The main results are summarized below. 

\begin{itemize}
  \item 
The detection rates of the \dcop emission in the 
prestellar and protostellar cores are higher than 
those of \n2dp, whereas \n2dp is more sensitive to 
the core evolution than \dcop in terms of the clear 
variations of $\sigma_{\rm obs}$. 
The commonly detected \dcop emission toward deeply embedded  
dense cores suggests that it is a good tracer of 
prestellar (detection rate 35\%)  and  early phase 
protostellar cores (detection rate 49\%). 
On the other hand, we find that \n2dp does not exclusively 
trace the prestellar cores and it is more frequently detected 
in the relatively earlier phase of the protostellar cores; the 
detection rate is 9\% for the prestellar cores and 38\% for 
the protostellar cores.  
This suggests that \n2dp is not the best tracer of prestellar 
cores at the sensitivity obtained in ASHES.

  \item 
Both \n2dp and \dcop abundances decrease with the 
core evolution. This is mainly caused by the effects 
of the H$_{2}$ density increasing rapidly and the 
temperature increasing slowly from the prestellar to 
protostellar phases in the identified cores.  
This can also explain that the \c18o abundance is 
higher in the prestellar cores than in the 
protostellar cores.

  \item 
The detection rate of the H$_{2}$CO emission toward dense cores 
is 52\%, three times higher than that of CH$_{3}$OH (17\%). 
The high detection rates of \h2co in both the prestellar 
(37\%) and protostellar cores (87\%) suggest that \h2co 
is commonly seen in the very early evolutionary phase. 
The line widths of \h2co are higher than those of 
\c18o, \n2dp, and \dcop toward the protostellar 
cores, which is likely due to the fact that the \h2co 
line is associated with more turbulent gas components 
related to protostellar activities (e.g., outflows). 
The \h2co abundances are found to increase with the 
evolution of the dense cores, as well as the line intensity 
and line width of \h2co ($3_{0,3}-2_{0,2}$) transition. 
These results indicate that \h2co could be used as a 
diagnostic tool to inferring star formation activities. 

\end{itemize}

\newpage

\acknowledgments
We thank the anonymous referee for constructive comments 
that helped improve this paper. 
This work is partly supported by the Korea Astronomy 
an Space Science Institute grant funded by the Korea 
government (MSIT) (Project No. 2022-1-840-05).
P.S. was partially supported by a Grant-in-Aid 
for Scientific Research (KAKENHI Number 18H01259 
and 22H01271). 
C.W.L. acknowledges the  support by the BasicScience 
Research Program through the National Research 
Foundation of Korea (NRF) funded by the Ministry of 
Education, Science and Technology 
(NRF-2019R1A2C1010851). 
SB is financially support by ANID Fondecyt Regular
(project \# 1220033) and the ANID BASAL projects 
ACE210002 and FB210003. 
K.T. was supported by JSPS KAKENHI (Grant Number 20H05645). 
This paper makes use of the following ALMA data: 
ADS/JAO.  ALMA\#2015.1.01539.S. 
ALMA is a partnership of ESO (representing its member states), 
NSF (USA) and NINS (Japan), together with NRC (Canada), MOST 
and ASIAA (Taiwan), and KASI (Republic of Korea), in cooperation 
with the Republic of Chile. The Joint ALMA Observatory is operated 
by ESO, AUI/NRAO and NAOJ.  Data analysis was in part carried out 
on the open use data analysis computer system at the Astronomy 
Data Center (ADC) of the National Astronomical Observatory of Japan.

\vspace{5mm}
\facilities{ALMA.}

\software{ CASA \citep{2007ASPC..376..127M}, 
APLpy \citep{2012ascl.soft08017R}, 
Astropy \citep{2013A&A...558A..33A}, 
Matplotlib \citep{4160265}, 
PySpecKit \citep{2022AJ....163..291G}.
  }


 \newpage 
\bibliography{chemistry}{}
\bibliographystyle{aasjournal}

\floattable
\begin{deluxetable*}{cccccccccc}
\tabletypesize{\scriptsize}
\tablecolumns{9}
\tablewidth{0pc}
\tablecaption{Summary of detected lines and their parameters. \label{tab:lines}}
\tablehead{
\colhead{Molecule} 	 &\colhead{Transition} &\colhead{Frequency} &\colhead{E$_{u}/k$} 
&\colhead{$S_{ij}\mu^{2}$} &\colhead{$n_{\rm crit}$} &\colhead{$Q_{\rm rot}$} &\colhead{Rotational Constants}
&\colhead{Beam Size} &\colhead{rms} \\
         		 & 	 &\colhead{(GHz)}	 &\colhead{(K)} 
&\colhead{D$^{2}$} &\colhead{(cm$^{-3}$)} & 				 &\colhead{(MHz)} 
&\colhead{(\arcsec)}  &\colhead{}
}
\startdata
DCO$^{+}$ & $3-2$ & 216.112 & 20.74 & 45.624 & 1.84E+06 & 0.58$T_{\rm ex}$ + 0.34 &  & 1.5 $\times$ 1.0 & 9.5$^{a}$	\\
C$_{2}$D & $3-2$ & 216.373 & 20.77 & 2.541 & 8.18E+05 & 3.47$T_{\rm ex}$ + 2.06 &  & 1.6 $\times$ 1.0 & 9.5$^{a}$	\\
SiO & $5-4$ & 217.105 & 14.48 & 48.146 & 1.22E+06 & $\frac{k_{\rm B}T_{\rm ex}}{hB_{0}} + \frac{1}{3}$ & $B_{0}$=21711.97 & 1.6 $\times$ 1.0 & 9.5$^{a}$	\\
DCN & $3-2$ & 217.238 & 20.85 & 80.501 & 2.16E+07 & 0.1$T_{\rm ex}^{3/2}$ + 50.51 &  & 1.7 $\times$ 1.0 & 9.5$^{a}$	\\
p-H$_{2}$CO & 3$_{0,3}-2_{0,2}$ & 218.222 & 20.95 & 16.308 & 2.56E+06 & $\frac{1}{3} \left( \frac{\pi k_{\rm B}^{3}T_{\rm ex}^{3} }{h^{3}A_{0}B_{0}C_{0} } \right)^{0.5}$ & $A_{0}$ = 281970.56 & 1.5 $\times$ 1.0 & 3.5$^{b}$	\\
p-H$_{2}$CO & 3$_{2,2}-2_{2,1}$ & 218.475 & 68.09 & 9.062 & 2.96E+06 &  & $B_{0}$ = 38833.987 & 1.5 $\times$ 1.0 & 3.5$^{b}$	\\
p-H$_{2}$CO & 3$_{2,1}-2_{2,0}$ & 218.76 & 68.11 & 9.062 & 3.36E+06 &  & $C_{0}$ = 34004.244 & 1.5 $\times$ 1.0 & 3.5$^{b}$	\\
CH$_{3}$OH & 4$_{2,2}-3_{1,2}$ & 218.44 & 45.46 & 13.906 & 2.04E+07 & $\frac{1}{2} \left( \frac{\pi k_{\rm B}^{3}T_{\rm ex}^{3} }{h^{3}A_{0}B_{0}C_{0} } \right)^{0.5}$ & $A_{0}$ = 127523.4 & 1.5 $\times$ 1.0 & 3.5$^{b}$	\\
 &  &  &  &  &  &  & $B_{0}$ = 24690.2 &  & 	\\
 &  &  &  &  &  &  & $C_{0}$ = 23759.7 &  & 	\\
C$^{18}$O & $2-1$ & 219.56 & 15.81 & 0.024 & 9.33E+03 & $\frac{k_{\rm B}T_{\rm ex}}{hB_{0}} + \frac{1}{3}$ & $B_{0}$=54891.42 & 1.5 $\times$ 1.0 & 3.5$^{b}$	\\
$^{13}$CS & $5-4$ & 231.22 & 32.73 & 38.335 & 4.46E+06 & 1.85$T_{\rm ex}$ - 3.32 &  & 1.4 $\times$ 1.0 & 9.5$^{a}$	\\
N$_{2}$D$^{+}$ & $3-2$ & 231.321 & 22.2 & 312.104 & 1.70E+06 & 4.87$T_{\rm ex}$ + 2.81 &  & 1.5 $\times$ 1.0 & 9.5$^{a}$	\\
\enddata
\tablenotetext{}{Notes. 
E$_{u}/k$ and $S_{ij}\mu^{2}$ are obtained from the 
Cologne Database for Molecular Spectroscopy  
\citep[CDMS\footnote{\url{https://cdms.astro.uni-koeln.de/cdms/portal/}};][]{2005JMoSt.742..215M}. 
The critical densities are estimated using the equation 
$n_{\rm crit} \, = \, A_{\rm u}/\gamma$, where Einstein A coefficients 
($A_{\rm u}$) and collisional rates ($\gamma$) at 20 K were obtained from the Leiden atomic and molecular database 
\citep{Schoier05}. 
For \n2dp, DCN, and \ccd we use the same collision rates 
as those for N$_{2}$H$^{+}$, HCN, and C$_{2}$H, respectively, 
since the lack of direct experimental constraints and the 
transitions between the the deuterated and non-deuterared 
isotopologue do not differ significantly. 
Collision rate value of $^{13}$CS from the its main 
isotopologue. 
Rotational constants (i.e., $A_{0}$, $B_{0}$, $C_{0}$) are retrieved  
from splatalogue database for astronomical 
spectroscopy\footnote{\url{https://splatalogue.online//}}. 
a: the unit is mJy beam$^{-1}$ per 0.17 km s$^{-1}$.
b: the unit is mJy beam$^{-1}$ per 1.3 km s$^{-1}$.
}
\end{deluxetable*}

\begin{deluxetable*}{ccccccc}
\tabletypesize{\scriptsize}
\tablecolumns{7}
\tablewidth{0pc}
\tablecaption{Detection rate 
\label{tab:detection}}
\tablehead{
\colhead{Molecules} &\colhead{Prestellar}	
&\colhead{}	 &\colhead{Protostellar} &\colhead{} &\colhead{}
&\colhead{All} \\
\cline{3-6}
\colhead{} &\colhead{}	
&\colhead{Outflow Core}	 &\colhead{Warm Core} &\colhead{Outflow\&Warm Core} &\colhead{Sum}
&\colhead{}\\ \hline
\colhead{Total Number} &\colhead{197}	
&\colhead{21}	 &\colhead{37} &\colhead{37} &\colhead{97}
&\colhead{294} 
}
\startdata
C$^{18}$O & 177 (89.8\%) & 18 (85.7\%) & 35 (94.6\%) & 37 (94.9\%) & 90 (92.8\%) & 267 (90.8\%)	\\
DCO$^{+}$ & 69 (35.0\%) & 8 (38.1\%) & 18 (48.6\%) & 21 (53.8\%) & 47 (48.5\%) & 116 (39.5\%)	\\
N$_{2}$D$^{+}$ & 17 (8.6\%) & 8 (38.1\%) & 7 (18.9\%) & 22 (56.4\%) & 37 (38.1\%) & 54 (18.4\%)	\\
H$_{2}$CO & 72 (36.5\%) & 14 (66.7\%) & 33 (89.2\%) & 37 (94.9\%) & 84 (86.6\%) & 156 (53.1\%)	\\
CH$_{3}$OH & - & - & 21 (56.8\%) & 30 (76.9\%) & 51 (52.6\%) & 51 (17.3\%)	\\
SiO & - & 3 (14.3\%) & - & 24 (61.5\%) & 27 (27.8\%) & 27 (9.2\%)	\\
$^{13}$CS & 2 (1.0\%) & 1 (4.8\%) & - & 1 (2.6\%) & 2 (2.1\%) & 4 (1.4\%)	\\
DCN & - & - & - & 7 (17.9\%) & 7 (7.2\%) & 7 (2.4\%)	\\
C$_{2}$D & 3 (1.5\%) & - & - & - & - & 3 (1.0\%)	\\
\enddata
\tablenotetext{}{
}
\end{deluxetable*}

\begin{table*}
\centering
\caption{Median and mean values of derived line parameters for each category}
\label{tab:mean}
\begin{tabular}{l c c c c c c c c c  c c c c}
\toprule
Name & 
\multicolumn{2}{c}{Prestellar} & 
\multicolumn{8}{c}{Protostellar} &
\multicolumn{2}{c}{All} \\ 
\cmidrule(l{2pt}r{2pt}){4-11}
	 & & &
\multicolumn{2}{c}{Outflow Core} &
\multicolumn{2}{c}{Warm Core} &
\multicolumn{2}{c}{Outflow\&Warm Core} &
\multicolumn{2}{c}{Sum} & \\
\cmidrule(l{2pt}r{2pt}){2-3} \cmidrule(l{2pt}r{2pt}){4-5} 
\cmidrule(l{2pt}r{2pt}){6-7} \cmidrule(l{2pt}r{2pt}){8-9}
\cmidrule(l{2pt}r{2pt}){10-11} \cmidrule(l{2pt}r{2pt}){12-13} 
& median & mean 
& median & mean 
& median & mean 
& median & mean 
& median & mean 
& median & mean &    \\   
	\midrule    
$I_{\rm C^{18}O}$ & 0.98 & 1.12 & 0.81 & 0.88 & 0.86 & 0.96 & 1.22 & 1.30 & 0.90 & 1.09 & 0.93 & 1.11	\\
$\sigma_{\rm C^{18}O}$ & 0.75 & 0.89 & 0.88 & 0.95 & 0.92 & 1.02 & 0.85 & 0.97 & 0.88 & 0.99 & 0.79 & 0.92	\\
$I_{\rm  N_{2}D^{+}}$ & 0.41 & 0.42 & 0.29 & 0.31 & 0.52 & 0.47 & 0.44 & 0.53 & 0.40 & 0.47 & 0.41 & 0.45	\\
$\sigma_{\rm  N_{2}D^{+}}$ & 0.22 & 0.21 & 0.21 & 0.25 & 0.19 & 0.21 & 0.24 & 0.26 & 0.23 & 0.25 & 0.22 & 0.23	\\
$I_{\rm DCO^{+}}$ & 0.32 & 0.37 & 0.37 & 0.46 & 0.33 & 0.37 & 0.36 & 0.37 & 0.35 & 0.39 & 0.33 & 0.38	\\
$\sigma_{\rm DCO^{+}}$ & 0.29 & 0.30 & 0.29 & 0.29 & 0.38 & 0.39 & 0.32 & 0.35 & 0.32 & 0.35 & 0.30 & 0.32	\\
$\sigma_{\rm CH_{3}OH}$ & - & - & - & - & 1.23 & 1.59 & 1.46 & 1.98 & 1.35 & 1.82 & 1.35 & 1.82	\\
$I_{\rm H_{2}CO}$ & 0.18 & 0.19 & 0.28 & 0.28 & 0.33 & 0.42 & 0.61 & 0.88 & 0.38 & 0.60 & 0.26 & 0.41	\\
$\sigma_{\rm H_{2}CO}$ & 0.72 & 0.84 & 0.96 & 1.06 & 1.19 & 1.36 & 1.85 & 2.21 & 1.33 & 1.68 & 1.00 & 1.29	\\
$I_{\rm ^{13}CS}$ & 0.38 & 0.38 & 0.27 & 0.27 & - & - & 0.18 & 0.18 & 0.22 & 0.22 & 0.28 & 0.30	\\
$\sigma_{\rm ^{13}CS}$ & 0.34 & 0.34 & 0.46 & 0.46 & - & - & 1.30 & 1.30 & 0.88 & 0.88 & 0.42 & 0.61	\\
$I_{\rm DCN}$ & - & - & - & - & - & - & 0.26 & 0.22 & 0.26 & 0.22 & 0.26 & 0.22	\\
$\sigma_{\rm DCN}$ & - & - & - & - & - & - & 1.21 & 1.04 & 1.21 & 1.04 & 1.21 & 1.04	\\
$I_{\rm C_{2}D}$ & 0.22 & 0.24 & - & - & - & - & - & - & - & - & 0.22 & 0.24	\\
$\sigma_{\rm C_{2}D}$ & 0.13 & 0.17 & - & - & - & - & - & - & - & - & 0.13 & 0.17	\\
\bottomrule
\end{tabular}
\tablenotetext{}{Notes. 
The units are K and \kms for peak intensity and observed linewidth, respectively.}
\end{table*}

\begin{table*}
\tiny
\centering
\caption{Median and mean values of derived core properties for each category}
\label{tab:Nline}
\begin{tabular}{l c c c c c c c c c  c c c}
\toprule
Name & 
\multicolumn{2}{c}{Prestellar} & 
\multicolumn{8}{c}{Protostellar} &
\multicolumn{2}{c}{All} \\ 
\cmidrule(l{2pt}r{2pt}){4-11}
	 & & &
\multicolumn{2}{c}{Outflow Core} &
\multicolumn{2}{c}{Warm Core} &
\multicolumn{2}{c}{Outflow\&Warm Core} &
\multicolumn{2}{c}{Sum} & \\
\cmidrule(l{1pt}r{2pt}){2-3} \cmidrule(l{2pt}r{2pt}){4-5} 
\cmidrule(l{1pt}r{2pt}){6-7} \cmidrule(l{2pt}r{2pt}){8-9}
\cmidrule(l{1pt}r{2pt}){10-11} \cmidrule(l{2pt}r{2pt}){12-13} 
& median & mean 
& median & mean 
& median & mean 
& median & mean 
& median & mean 
& median & mean     \\   
	\midrule    
$T_{\rm NH_{3}}$(K) & 14.1 & 14.2 & 12.8 & 13.8 & 14.8 & 14.3 & 15.3 & 14.9 & 14.8 & 14.4 & 14.4 & 14.3	\\ \hline 
 &  & 						\multicolumn{7}{c}{Core averaged volume densities (cm$^{-3}$) and averaged column densities (cm$^{-2}$)}				 &  &  &  & 			\\ \hline
$n_{\rm H_2}$ & 1.06E+06 & 1.51E+06 & 1.52E+06 & 2.50E+06 & 1.20E+06 & 2.69E+06 & 3.42E+06 & 5.28E+06 & 2.30E+06 & 3.69E+06 & 1.25E+06 & 2.23E+06	\\
$N_{\rm H_2}$ & 2.10E+22 & 2.56E+22 & 3.25E+22 & 3.83E+22 & 2.57E+22 & 4.15E+22 & 5.64E+22 & 7.91E+22 & 4.36E+22 & 5.59E+22 & 2.56E+22 & 3.56E+22	\\ \hline 
 &  &  &  & 		\multicolumn{5}{c}{Column densities (cm$^{-2}$)}				 &  &  &  & 			\\ \hline
$N_{\rm C^{18}O}$ & 1.10E+15 & 1.42E+15 & 8.35E+14 & 1.30E+15 & 1.18E+15 & 1.54E+15 & 1.79E+15 & 2.20E+15 & 1.29E+15 & 1.77E+15 & 1.16E+15 & 1.54E+15	\\
$N_{\rm N_{2}D^{+}}$ & 2.63E+11 & 3.30E+11 & 2.28E+11 & 2.61E+11 & 3.55E+11 & 4.06E+11 & 2.96E+11 & 4.39E+11 & 2.74E+11 & 3.94E+11 & 2.72E+11 & 3.74E+11	\\
$N_{\rm DCO^{+}}$ & 1.47E+11 & 1.89E+11 & 1.71E+11 & 2.12E+11 & 2.22E+11 & 2.37E+11 & 1.92E+11 & 2.05E+11 & 1.92E+11 & 2.18E+11 & 1.67E+11 & 2.01E+11	\\
$N_{\rm CH_{3}OH}$ & - & - & - & - & 8.36E+12 & 2.09E+13 & 1.86E+13 & 3.61E+13 & 1.74E+13 & 2.98E+13 & 1.74E+13 & 2.98E+13	\\
$N_{\rm H_{2}CO}$ & 1.39E+12 & 1.61E+12 & 2.84E+12 & 3.00E+12 & 4.29E+12 & 5.83E+12 & 1.20E+13 & 2.04E+13 & 6.11E+12 & 1.18E+13 & 2.43E+12 & 7.09E+12	\\
$N_{\rm ^{13}CS}$ & 1.14E+12 & 1.14E+12 & 1.06E+12 & 1.06E+12 & - & - & 1.89E+12 & 1.89E+12 & 1.47E+12 & 1.47E+12 & 1.29E+12 & 1.31E+12	\\
$N_{\rm DCN}$ & - & - & - & - & - & - & 7.73E+11 & 1.05E+12 & 7.73E+11 & 1.05E+12 & 7.73E+11 & 1.05E+12	\\
$N_{\rm C_{2}D}$ & 8.91E+12 & 8.33E+12 & - & - & - & - & - & - & - & - & 8.91E+12 & 8.33E+12	\\
$N_{\rm SiO}$ & - & - & 1.74E+12 & 1.46E+12 & - & - & 1.70E+12 & 5.59E+12 & 1.74E+12 & 5.13E+12 & 1.74E+12 & 5.13E+12	\\ \hline
 &  &  &  &  & \multicolumn{3}{c}{Molecular abundances}		 &  &  &  &  & 			\\ \hline
$X_{\rm C^{18}O}$ & 4.95E-08 & 6.74E-08 & 3.53E-08 & 3.68E-08 & 3.85E-08 & 6.37E-08 & 2.33E-08 & 2.95E-08 & 2.99E-08 & 4.43E-08 & 4.23E-08 & 5.96E-08	\\
$X_{\rm N_{2}D^{+}}$ & 7.57E-12 & 8.07E-12 & 5.94E-12 & 5.99E-12 & 7.78E-12 & 7.91E-12 & 4.90E-12 & 5.07E-12 & 5.23E-12 & 5.81E-12 & 6.12E-12 & 6.52E-12	\\
$X_{\rm DCO^{+}}$ & 5.23E-12 & 5.95E-12 & 4.79E-12 & 4.71E-12 & 6.58E-12 & 7.60E-12 & 2.46E-12 & 2.84E-12 & 4.02E-12 & 4.98E-12 & 4.78E-12 & 5.56E-12	\\
$X_{\rm CH_{3}OH}$ & - & - & - & - & 2.75E-10 & 4.44E-10 & 3.00E-10 & 4.72E-10 & 2.83E-10 & 4.60E-10 & 2.83E-10 & 4.60E-10	\\
$X_{\rm H_{2}CO}$ & 5.20E-11 & 6.88E-11 & 8.09E-11 & 8.74E-11 & 1.59E-10 & 2.40E-10 & 1.76E-10 & 2.80E-10 & 1.53E-10 & 2.32E-10 & 8.87E-11 & 1.57E-10	\\
$X_{\rm ^{13}CS}$ & 6.03E-11 & 6.03E-11 & 7.94E-11 & 7.94E-11 & - & - & 1.52E-11 & 1.52E-11 & 4.73E-11 & 4.73E-11 & 5.41E-11 & 5.38E-11	\\
$X_{\rm DCN}$ & - & - & - & - & - & - & 8.30E-12 & 8.55E-12 & 8.30E-12 & 8.55E-12 & 8.30E-12 & 8.55E-12	\\
$X_{\rm C_{2}D}$ & 1.79E-10 & 2.50E-10 & - & - & - & - & - & - & - & - & 1.79E-10 & 2.50E-10	\\
$X_{\rm SiO}$ & - & - & 4.05E-11 & 4.38E-11 & - & - & 3.26E-11 & 5.69E-11 & 3.35E-11 & 5.54E-11 & 3.35E-11 & 5.54E-11	\\
\bottomrule
\end{tabular}
\end{table*}

\clearpage
\appendix

\section{Additional Figures and Table}
\label{app:fig}
\begin{figure*}[!ht]
\centering
\includegraphics[scale=0.24]{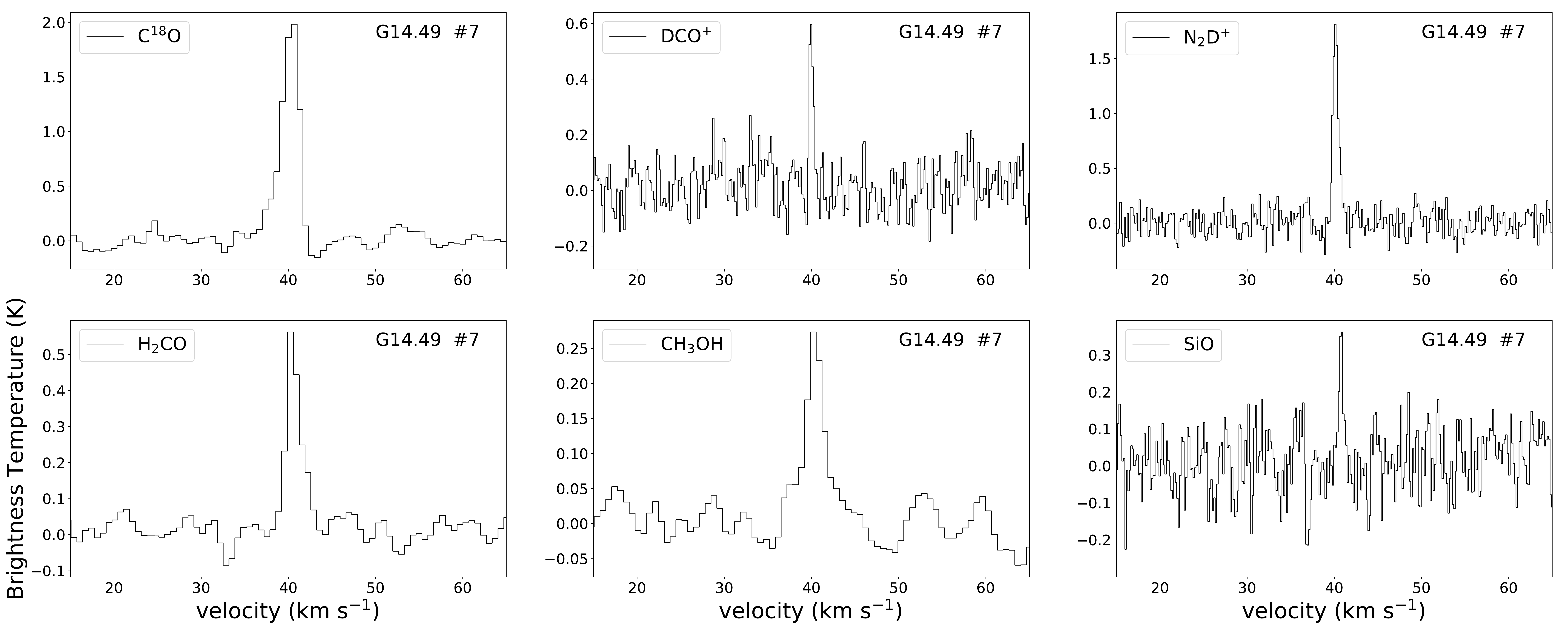}
\caption{
Examples of core-averaged spectra of \c18o, \dcop, \n2dp, 
\h2co, \ch3oh, and SiO for core \#7 in G14.49. 
}
 \label{fig:spec}
\end{figure*}

\begin{figure*}[!ht]
\centering
\includegraphics[scale=0.24]{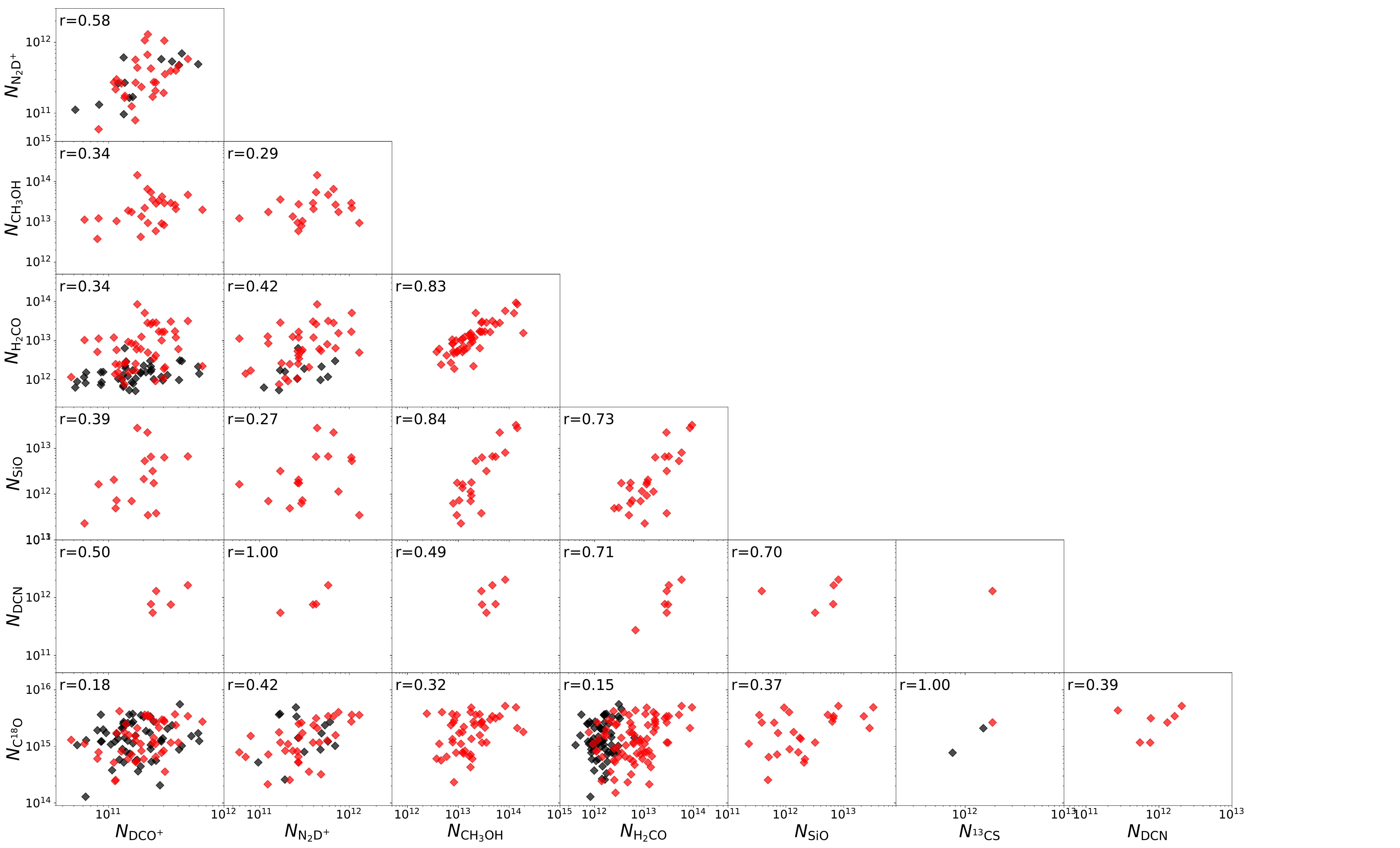}
\caption{
Column densities of each molecule plotted against one 
another. 
The black and red points represent the prestellar and 
protostellar cores, respectively. 
}
 \label{fig:Nmol}
\end{figure*}

Figure~\ref{fig:spec} shows the core-averaged spectra 
of the detected molecular lines in G14.49 as examples.   
Figures~\ref{fig:comb}-\ref{fig:comb4} shows velocity-integrated intensity 
maps of \n2dp, \dcop, DCN, \h2co, \ch3oh, and \c18o lines 
emission for each clump. 
Figure~\ref{fig:Nmol} shows the correlation of each pair of 
molecular column densities.  
Table~\ref{tab:linepara} summaries the derived parameters 
of detected lines.

\begin{figure*}
\center
\includegraphics[scale=0.34]{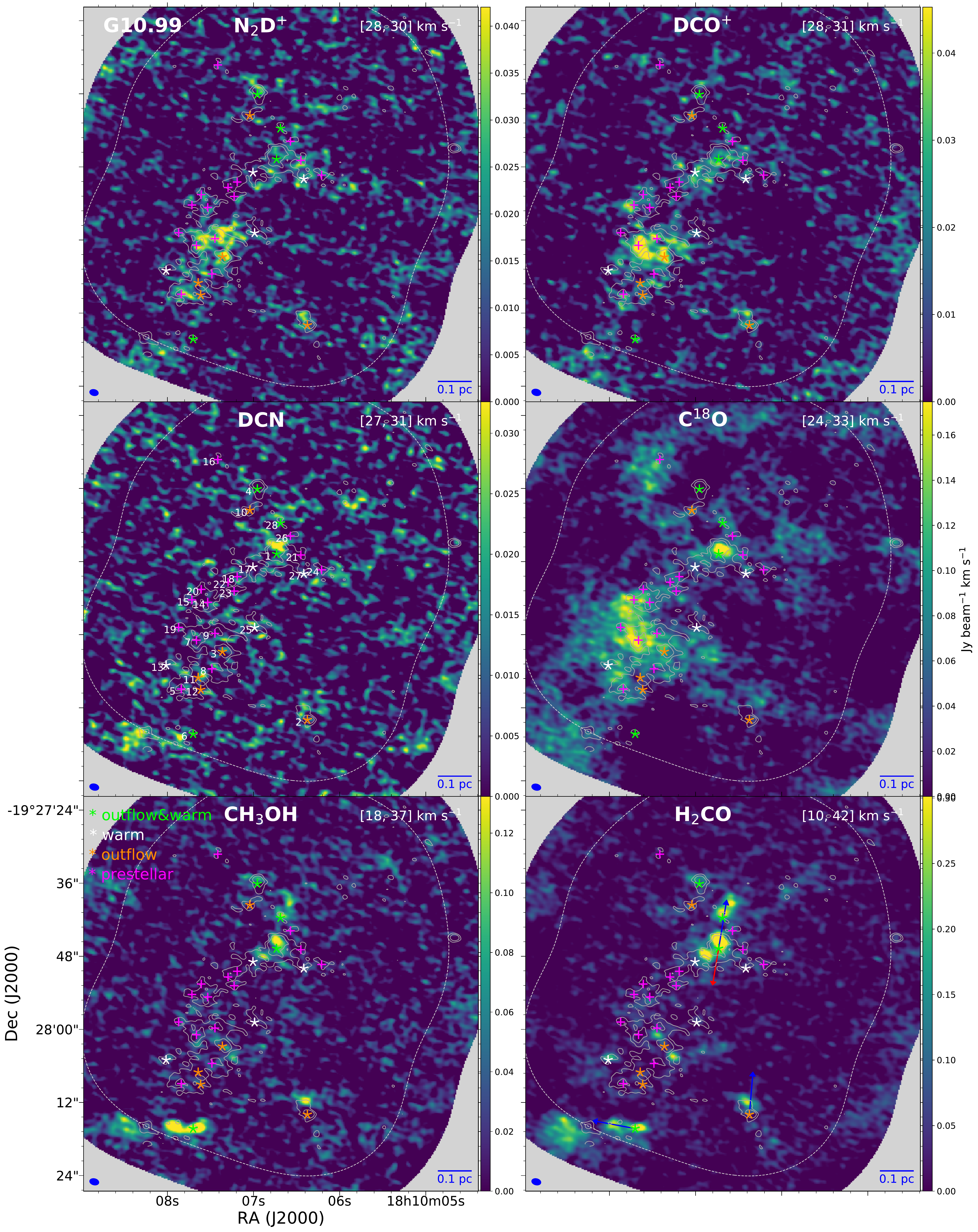}
\caption{The velocity-integrated intensity maps of \n2dp, 
\dcop, DCN, \c18o , \ch3oh, and \h2co emission lines  
toward G10.99. 
Fuchsia plusses, yellow, white, and Green asterisks symbols 
indicate prestellar candidates (category 1), outflow cores 
(category 2), warm cores (category 3), and outflow+warm 
cores (category 4). 
The white dashed line shows 30\% of the sensitivity 
level of the mosaic in the ALMA continuum image. 
The blue and red arrows in middle right panel indicate  
directions of the blueshifted and redshifted CO outflow lobes   
(see Paper \2), respectively. 
The gray contours in each panel shows the 
1.3 mm continuum emission.  The contour levels 
are (3, 6) $\times \, \sigma$ , 
with $\sigma$ = 0.115 mJy beam$^{-1}$. 
The beam size is shown in the lower left corner of 
each panel.}
\label{fig:comb}
\end{figure*}

\begin{figure*}\ContinuedFloat 
\center
\includegraphics[scale=0.36]{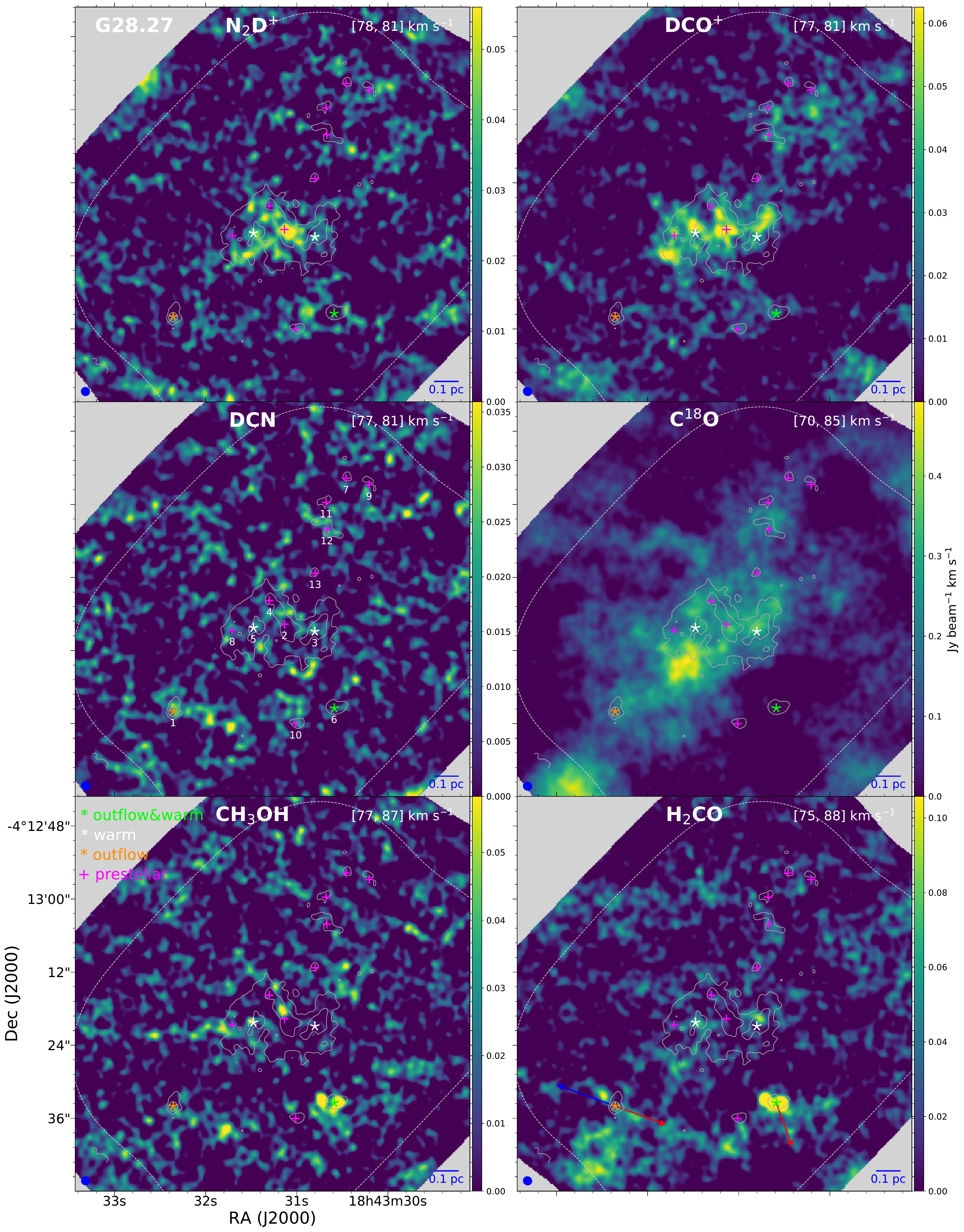}
\caption{}
\end{figure*}

\begin{figure*}\ContinuedFloat 
\center
\includegraphics[scale=0.36]{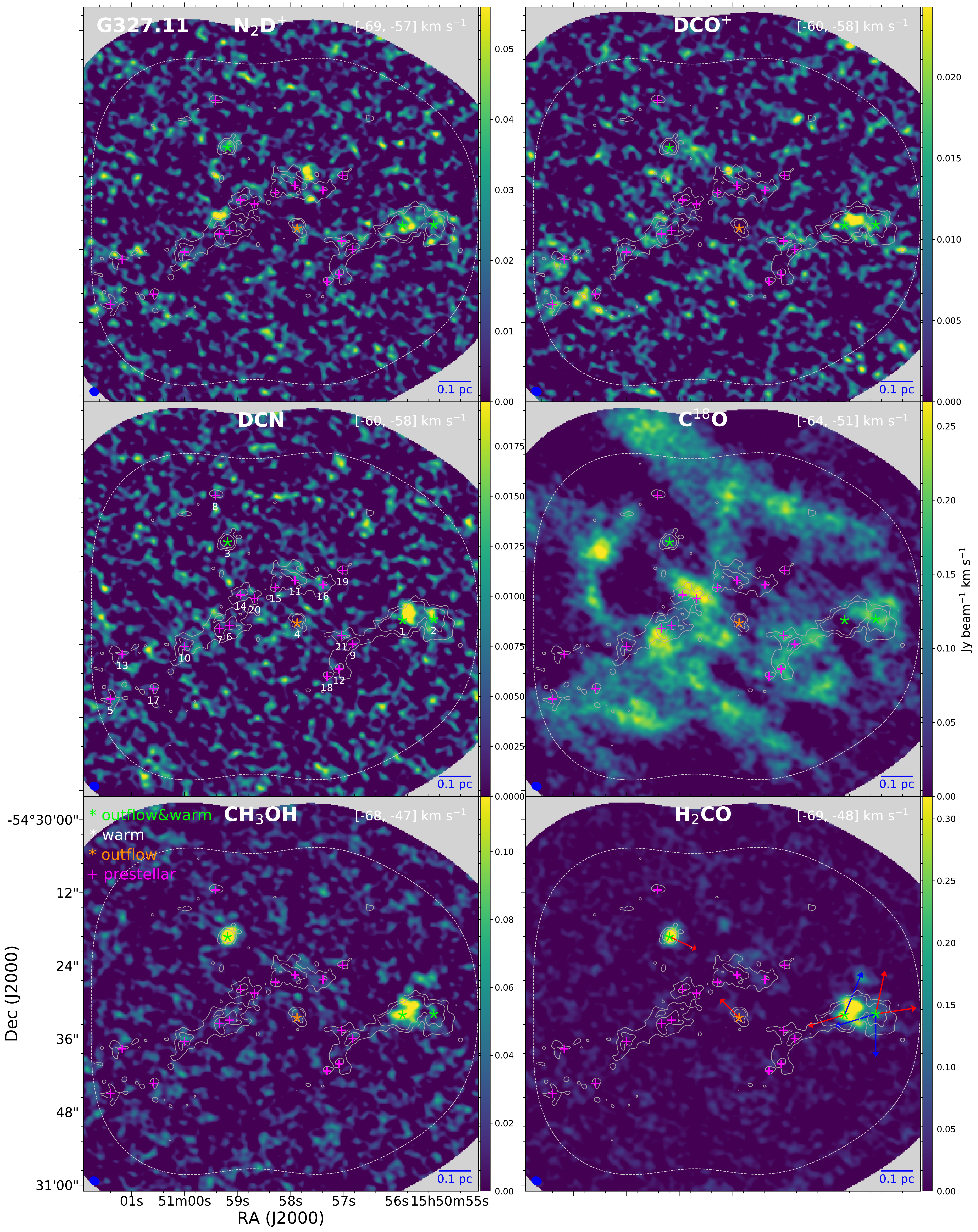}
\caption{}
\end{figure*}

\begin{figure*} 
\center
\includegraphics[scale=0.36]{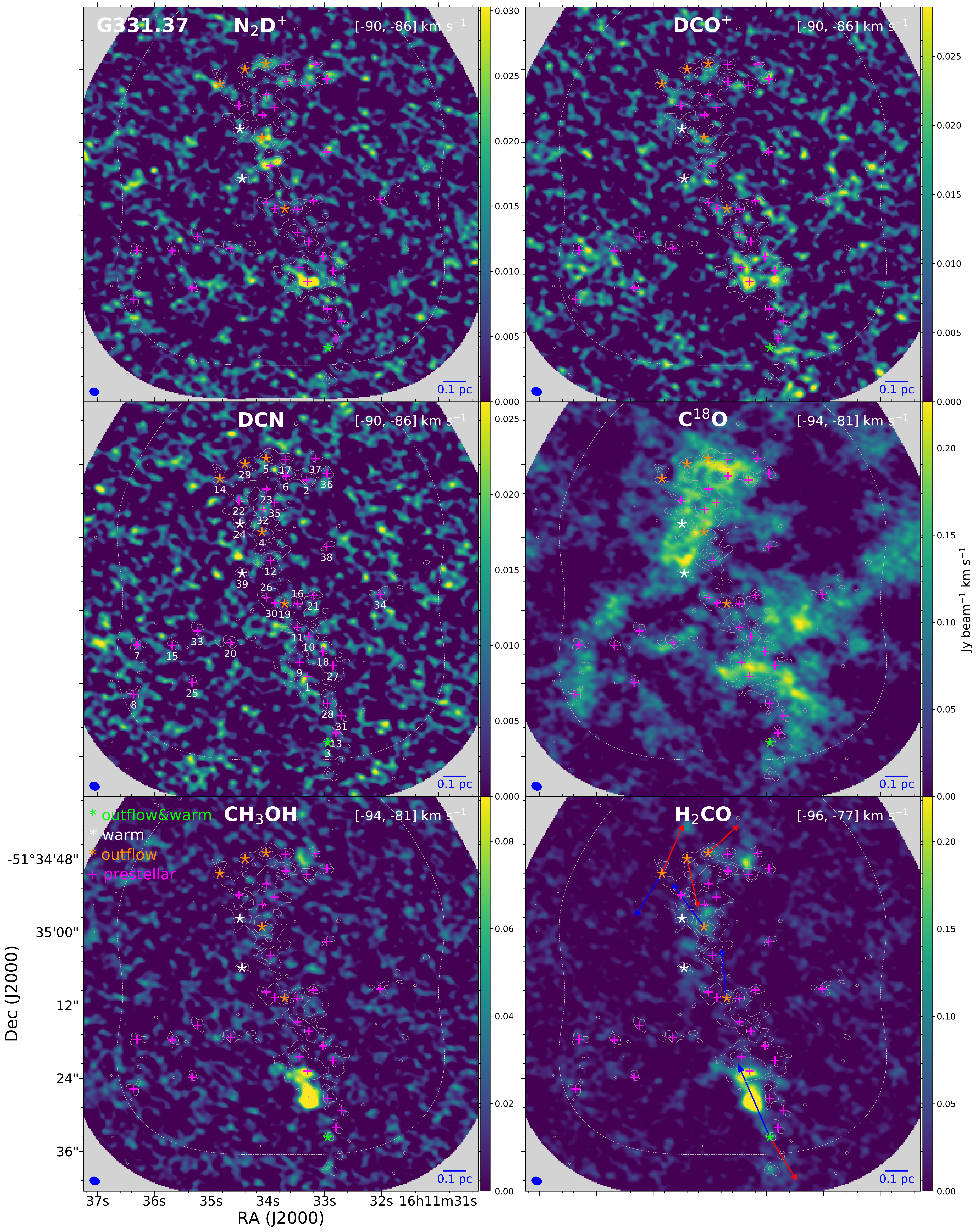}
\caption{same as Figure~\ref{fig:comb}, but for different sources.}
\label{fig:comb1}
\end{figure*}

\begin{figure*}\ContinuedFloat 
\center
\includegraphics[scale=0.36]{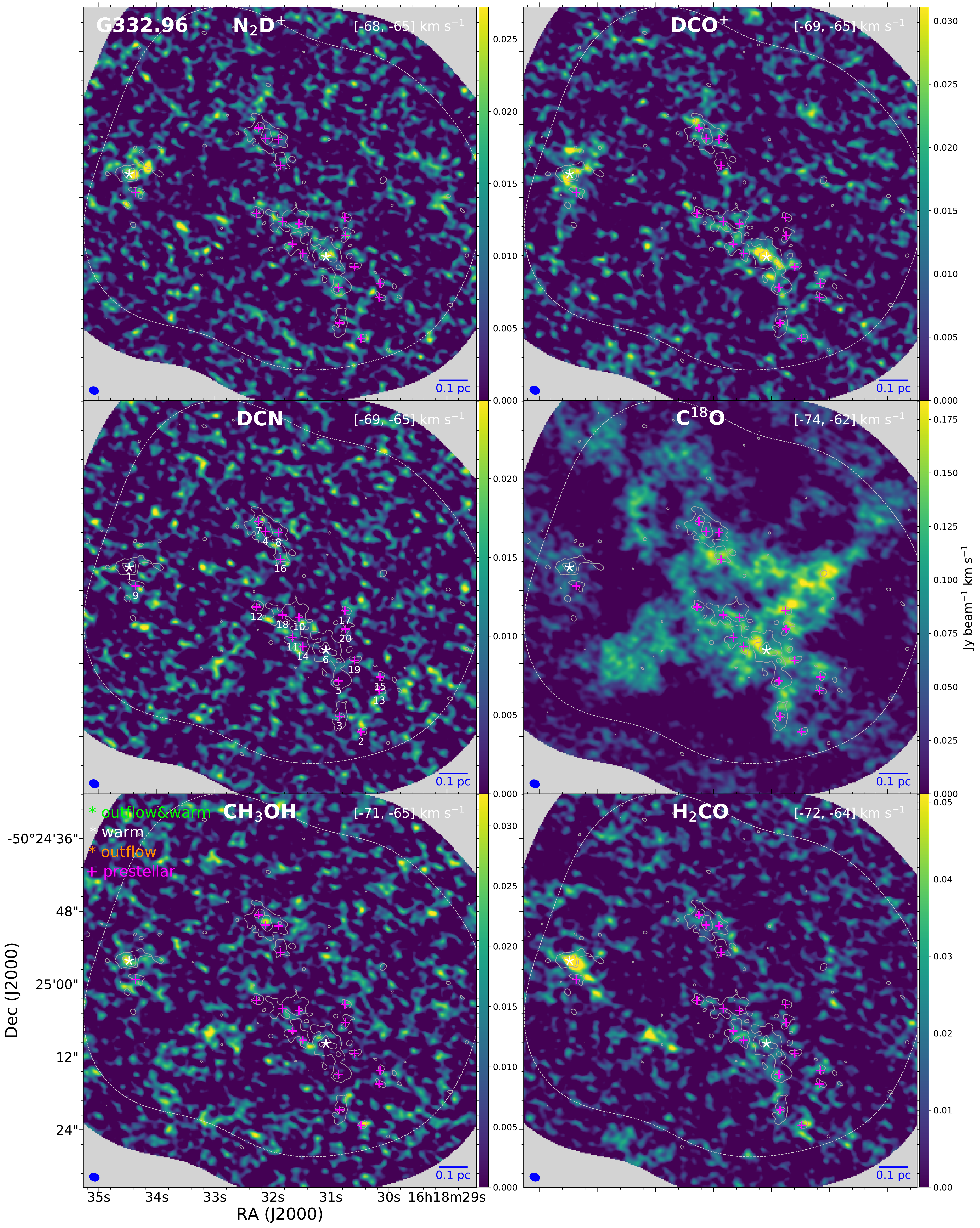}
\caption{}
\end{figure*}

\begin{figure*}\ContinuedFloat 
\center
\includegraphics[scale=0.36]{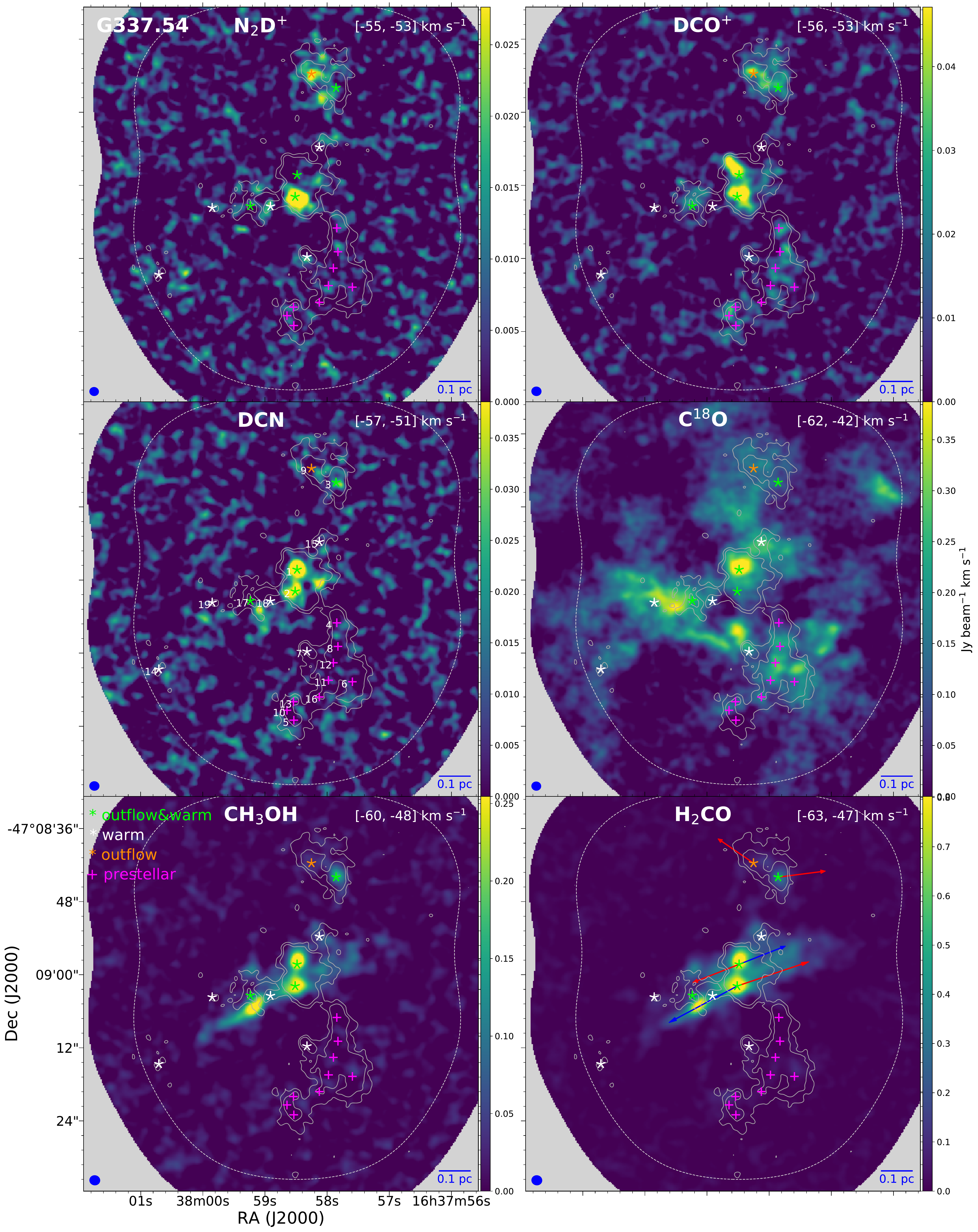}
\caption{}
\end{figure*}

\begin{figure*}
\center
\includegraphics[scale=0.36]{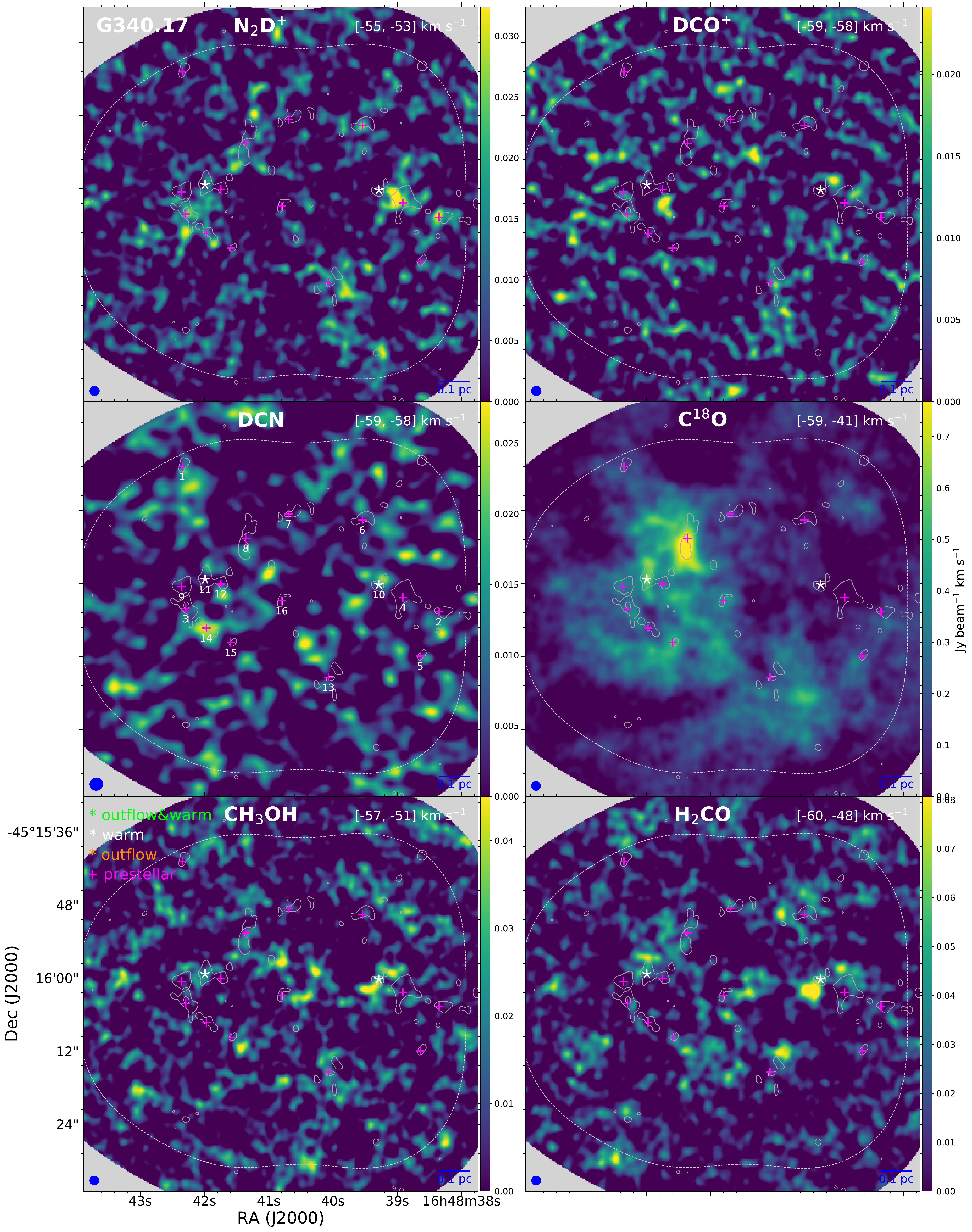}
\caption{same as Figure~\ref{fig:comb}, but for different sources.}
\label{fig:comb2}
\end{figure*}

\begin{figure*}
\center
\includegraphics[scale=0.36]{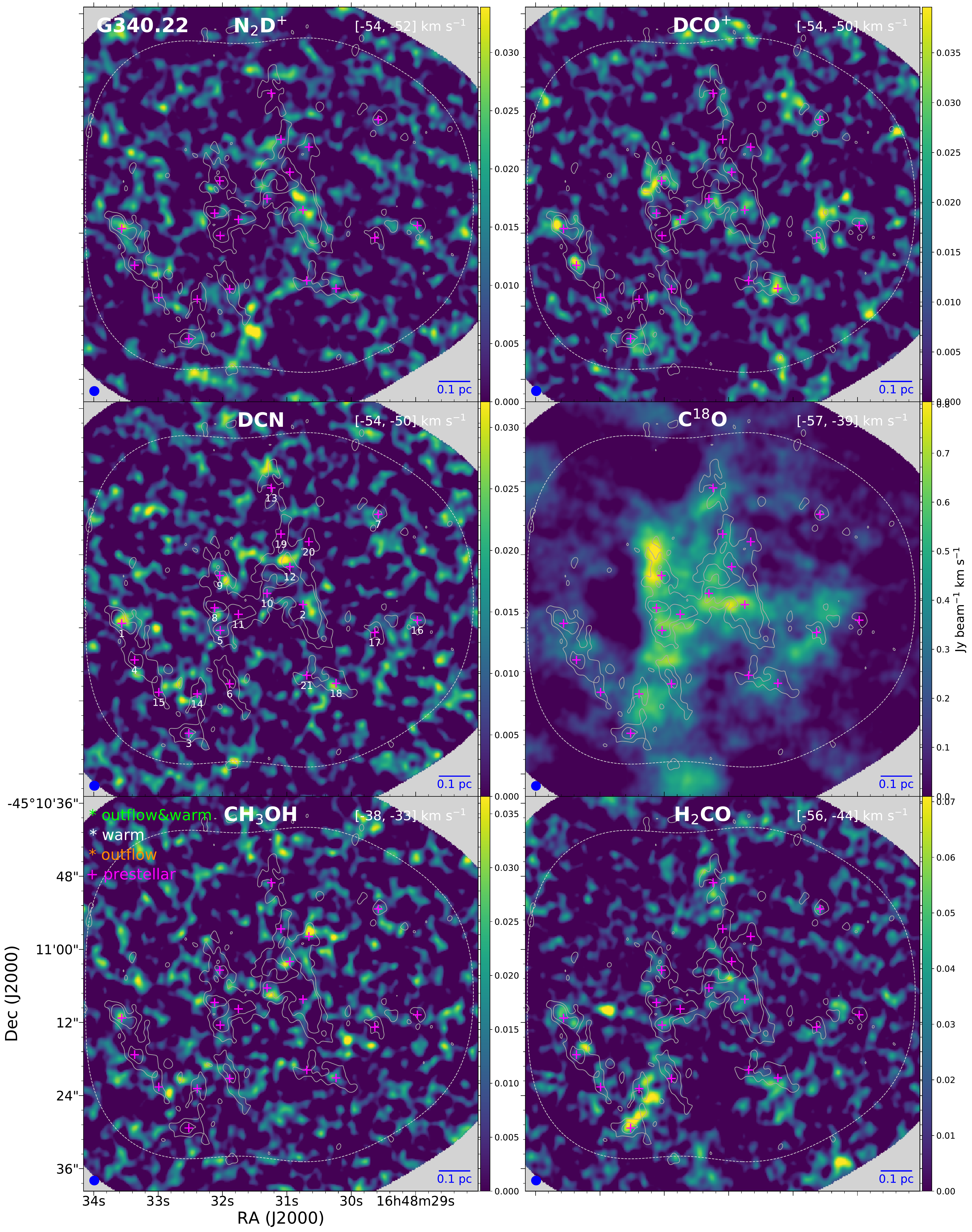}
\caption{same as Figure~\ref{fig:comb}, but for different sources.}
\label{fig:comb3}
\end{figure*}

\begin{figure*}\ContinuedFloat 
\center
\includegraphics[scale=0.36]{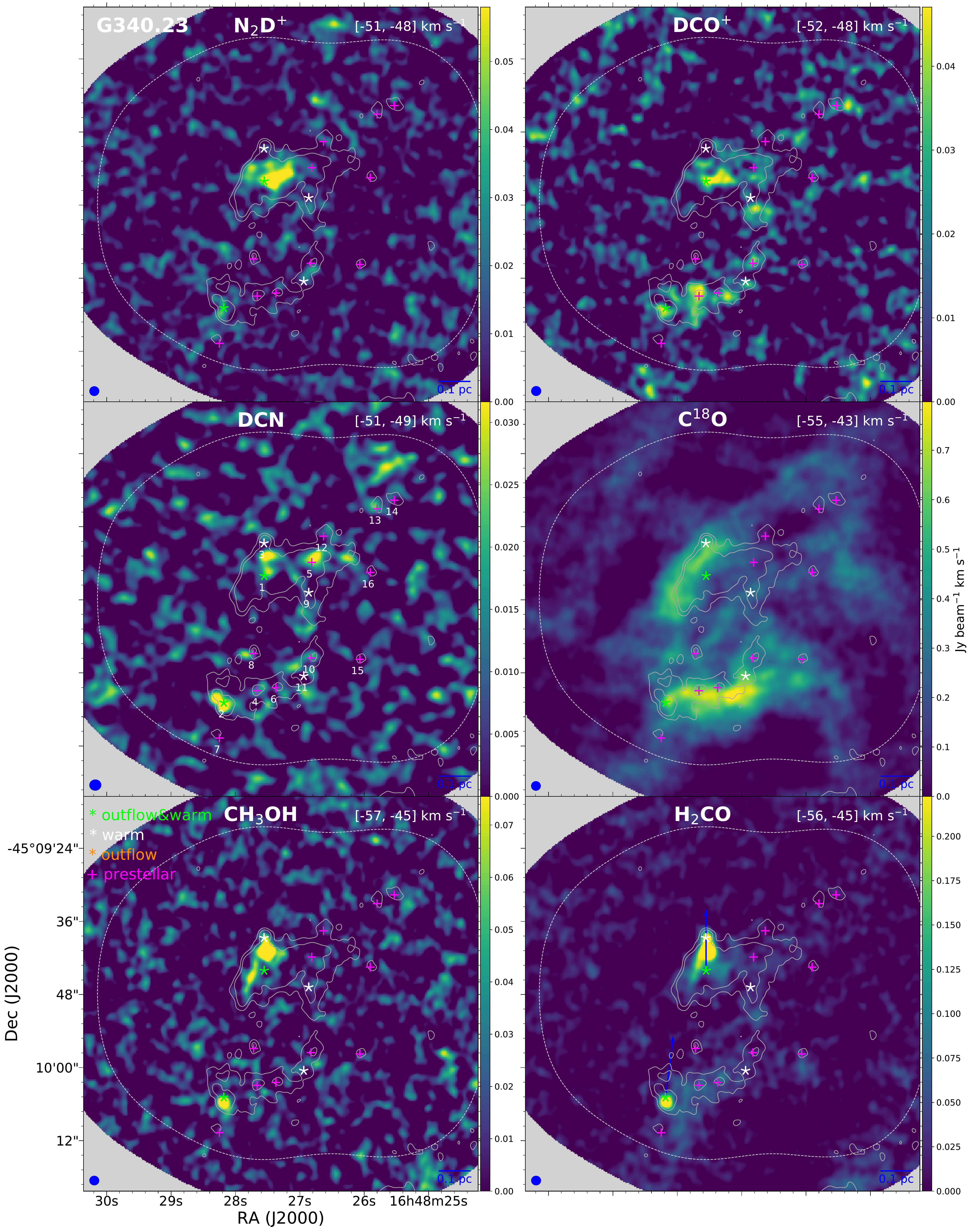}
\caption{}
\end{figure*}

\begin{figure*}\ContinuedFloat 
\center
\includegraphics[scale=0.36]{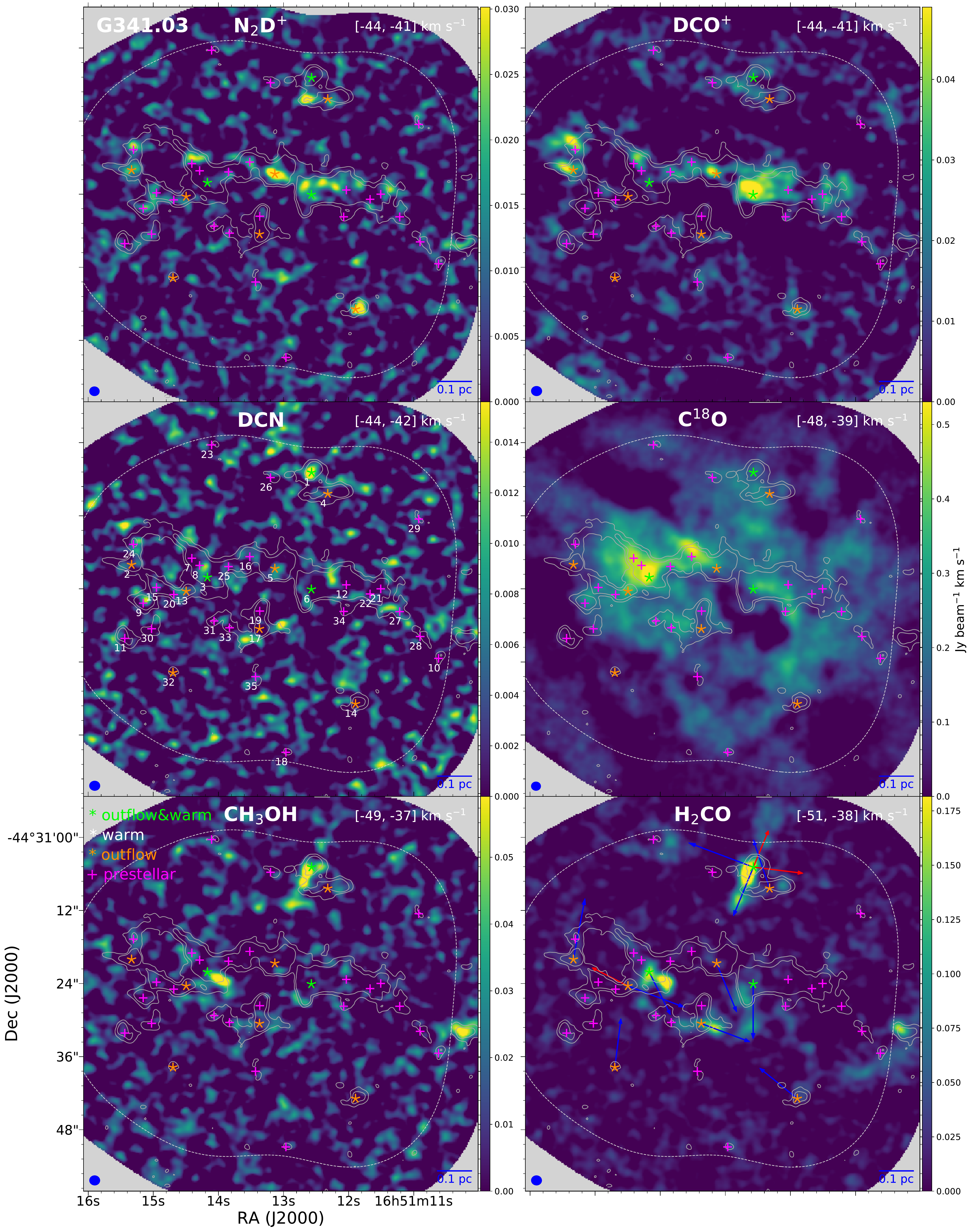}
\caption{}
\end{figure*}

\begin{figure*}
\center
\includegraphics[scale=0.36]{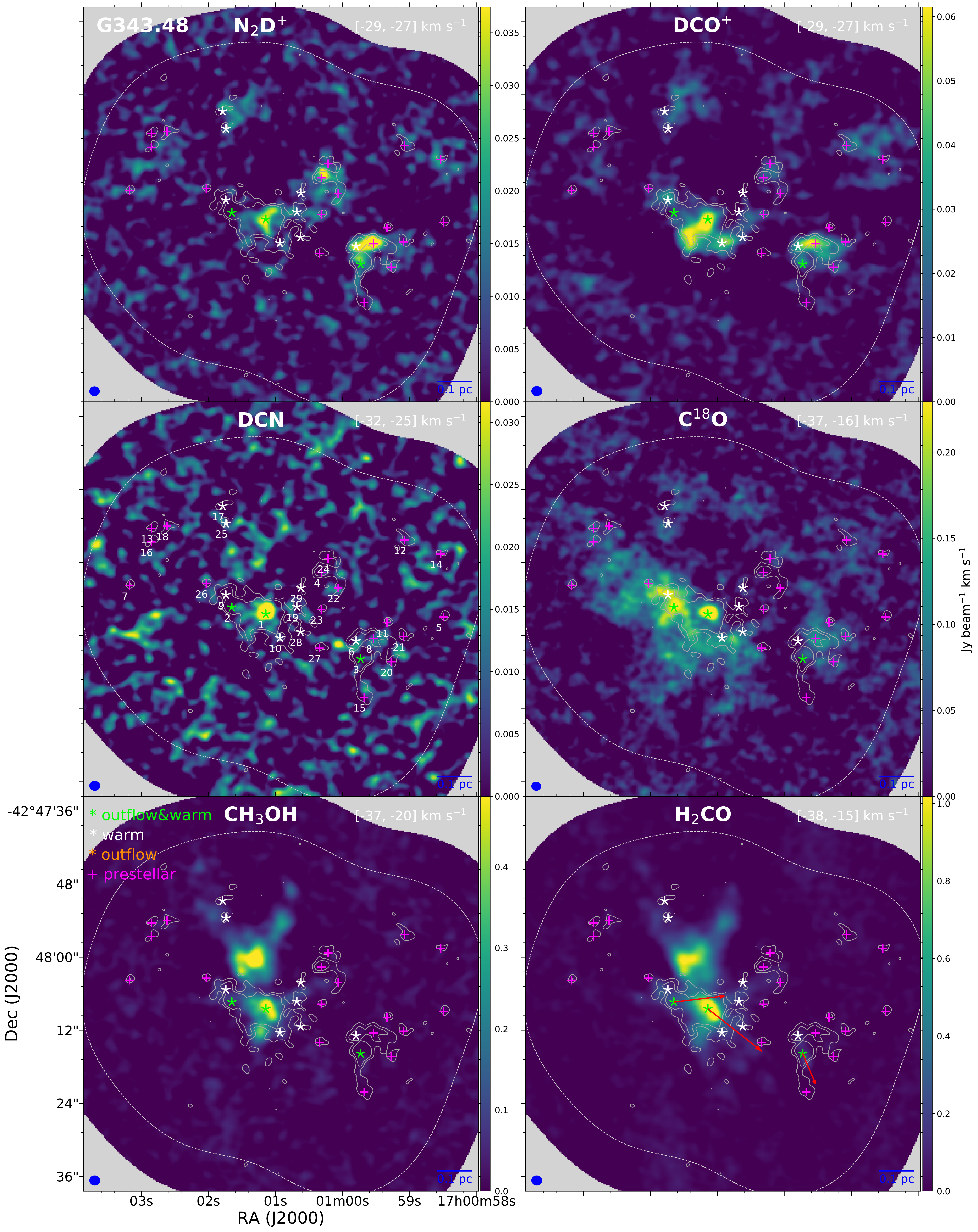}
\caption{same as Figure~\ref{fig:comb}, but for different sources.}
\label{fig:comb4}
\end{figure*}

\floattable
\begin{deluxetable*}{ccccccccccccc}
\tabletypesize{\scriptsize}
\tablewidth{0pc}
\tablecaption{Summary of derived parameters of detected lines. \label{tab:linepara}}
\tablehead{
\colhead{Molecule} & \colhead{core	}&  \colhead{$N_{\rm H_2}$} & \colhead{$n_{\rm H_2}$}& \colhead{$T_{\rm NH_3}$} & \colhead{$I_{\rm C^{18}O}$} & \colhead{$v_{\rm C^{18}O}$} & \colhead{$\sigma_{\rm C^{18}O}$} & \colhead{$N_{\rm C^{18}O}$}  & \colhead{$I_{\rm DCO^{+}}$} & \colhead{$v_{\rm DCO^{+}}$} &
\colhead{...} 
\\
 & &  \colhead{$\times 10^{22}$ cm$^{-2}$} & \colhead{$\times 10^{5}$ cm$^{-3}$} & \colhead{K} & \colhead{K} & \colhead{km s$^{-1}$} & \colhead{km s$^{-1}$} &
\colhead{$\times 10^{14}$ cm$^{-2}$	} & \colhead{K} & \colhead{km s$^{-1}$} &
...
}
\startdata
G10.99 & 1 & 4.78 & 1.59 & 13.4 & 0.69 (0.09) & 0.69 (0.09) & 30.16 (0.12) & 0.76 (0.12) & 0.41 (0.03) & 29.75 (0.03) & ...	\\ 
G10.99 & 2 & 8.31 & 7.80 & 12.5 & 0.39 (0.08) & 0.39 (0.08) & 29.96 (0.13) & 0.49 (0.11) & - & - & ...	\\ 
G10.99 & 3 & 7.80 & 5.21 & 11.7 & 0.65 (0.04) & 0.65 (0.04) & 29.92 (0.07) & 1.07 (0.07) & 0.51 (0.05) & 29.63 (0.05) & ...	\\ 
G10.99 & 4 & 4.41 & 2.80 & 13.0 & - & - & - & - & - & - & ...	\\ 
G10.99 & 5 & 5.55 & 2.33 & 13.3 & 0.62 (0.04) & 0.62 (0.04) & 29.28 (0.09) & 1.12 (0.09) & 0.79 (0.11) & 29.90 (0.03) & ...	\\ 
G10.99 & 6 & 5.55 & 5.51 & 13.0 & - & - & - & - & - & - & ...	\\ 
G10.99 & 7 & 5.29 & 2.15 & 12.3 & 1.23 (0.05) & 1.23 (0.05) & 29.62 (0.04) & 0.85 (0.04) & 1.74 (0.07) & 29.56 (0.01) & ...	\\ 
G10.99 & 8 & 4.92 & 1.59 & 12.2 & 0.80 (0.04) & 0.80 (0.04) & 30.13 (0.04) & 0.70 (0.04) & 0.44 (0.06) & 29.83 (0.05) & ...	\\ 
G10.99 & 9 & 4.25 & 2.43 & 13.3 & 0.84 (0.05) & 0.84 (0.05) & 29.38 (0.05) & 0.74 (0.05) & 0.66 (0.07) & 29.52 (0.04) & ...	\\ 
G10.99 & 10 & 3.32 & 1.53 & 11.7 & - & - & - & - & 0.29 (0.05) & 30.46 (0.05) & ...	\\ 
G10.99 & 11 & 5.56 & 6.16 & 11.4 & 0.33 (0.06) & 0.33 (0.06) & 30.74 (0.23) & 1.10 (0.23) & 0.87 (0.16) & 29.78 (0.03) & ...	\\ 
G10.99 & 12 & 4.85 & 4.61 & 11.3 & 0.53 (0.09) & 0.53 (0.09) & 30.14 (0.11) & 0.54 (0.11) & - & - & ...	\\ 
G10.99 & 13 & 3.21 & 1.55 & 13.0 & 0.80 (0.06) & 0.80 (0.06) & 28.99 (0.08) & 0.96 (0.08) & - & - & ...	\\ 
G10.99 & 14 & 3.29 & 2.22 & 13.3 & 0.63 (0.07) & 0.63 (0.07) & 29.50 (0.13) & 1.03 (0.13) & 0.32 (0.06) & 28.76 (0.08) & ...	\\ 
G10.99 & 15 & 3.38 & 1.66 & 13.3 & 1.10 (0.09) & 1.10 (0.09) & 29.54 (0.08) & 0.91 (0.08) & - & - & ...	\\ 
... & ... & ... & ... & ... & ...			 & ...			 & ...			 & ...			 & ...			 & 	...		 & ...	\\ 
G343.48 & 23 & 1.06 & 1.35 & 12.9 & - & - & - & - & 0.38 (0.06) & -28.62 (0.02) & ...	\\ 
G343.48 & 24 & 1.10 & 0.44 & 13.6 & - & - & - & - & 0.23 (0.04) & -28.93 (0.04) & ...	\\ 
G343.48 & 25 & 0.84 & 0.47 & 15.2 & 0.18 (0.02) & -28.25 (0.15) & 0.92 (0.15) & 2.32 (0.51) & - & - & ...	\\ 
G343.48 & 26 & 0.97 & 1.10 & 13.6 & 1.00 (0.03) & -28.84 (0.02) & 0.60 (0.02) & 8.78 (0.62) & - & - & ...	\\ 
G343.48 & 27 & 0.73 & 0.58 & 16.3 & 0.46 (0.03) & -28.73 (0.04) & 0.40 (0.02) & 2.63 (0.25) & - & - & ...	\\ 
G343.48 & 28 & 0.77 & 0.61 & 16.4 & 0.81 (0.03) & -28.40 (0.02) & 0.52 (0.02) & 6.05 (0.47) & 0.33 (0.05) & -28.38 (0.03) & ...	\\ 
G343.48 & 29 & 1.05 & 0.62 & 12.0 & 0.11 (0.02) & -26.20 (0.17) & 0.88 (0.17) & 1.52 (0.40) & - & - & ...	\\ 
\enddata
\tablenotetext{}{Notes. This table contains the  
column density, volume density, $T_{\rm NH_3}$, 
and derived parameters of lines for each dense core. 
(This table is available in its entirety in 
machine-readable form.)
}
\end{deluxetable*}

\clearpage

\section{Ammonia excitation temperature}
\label{app:tnh3}
Excitation temperatures were obtained using the method described in \cite{2015PASP..127..266M}. The ammonia (1,1) and (2,2) transition spectra were modelled using five-component Gaussian models with seven parameters (systemic velocity, line width and five amplitude parameters, one for each hyperfine component).  The best-fit model parameters were obtained using scipy's ``curve\_fit'' routine, using the TRF (``Trust Region Reflective'') algorithm.

From these parameters, the optical depth for the (1,1) transition, $\tau(1,1)$, was calculated from the ratio of the brightness of the satellite hyperfine transitions to the main component; this was solved numerically, using scipy's ``root'' routine, using the Levenberg-Marquardt algorithm to minimise the sum of squares error of the four ratios simultaneously.

Finally, the brightness temperatures, $T_{\rm B}(1,1)$ and $T_{\rm B}(2,2)$, and the optical depth were combined with the line width parameter ($\Delta v(1,1)$ and $\Delta v(2,2)$, common to all hyperfine components) to calculate the excitation temperature, $T_{\rm NH_3}$, as
 \begin{equation}
 T_{\rm NH_3} = \frac{-41.5}
 {\ln\left[
       -\frac{0.283 \Delta v(2,2)}{\tau(1,1)\Delta v(1,1)}
       \ln\left(
           1- \frac{T_B(2,2)}{T_B(1,1)} (1-\exp[-\tau(1,1)] )
          \right)
     \right]
 }.
\end{equation}
Full details of this method for calculating optical depth and excitation temperature are given in \cite{2015PASP..127..266M}.

The error bounds for the excitation temperature were then computed using a Monte Carlo approach.  For each model parameter, a randomised value was repeatedly drawn (1000 times) from a Gaussian distribution with mean equal to the optimal parameter value and variance equal to the variance of the parameter estimates, as reported by the fitting routine.  The optical depth and excitation temperature were then calculated using each set of randomised parameters, and the variance of the resulting distribution calculated to yield the temperature uncertainty.

\clearpage

\section{Column density}
\label{app:column}
Assuming local thermodynamic equilibrium (LTE)  and thin optical 
depths in the molecular line, the column densities of molecule can 
be calculated  following 
\citep{2015PASP..127..266M}\footnote{\url{https://github.com/ShanghuoLi/calcu}}
\begin{equation}
\label{N_mol}
N_{\rm mol} = \frac{3h}{8 \pi^{3} R}  
\frac{Q_{\rm rot}}{S \mu^{2} g_{\rm u}} 
\frac{{\rm exp}(E_{\rm u}/k T_{\rm ex})}{{\rm exp}(h\nu/k T_{\rm ex}) \, - \,1}
\left(J_{\nu}(T_{\rm ex}) \, - \, J_{\nu}(T_{\rm bg}) \right) ^{-1}
\int\frac{T_{\rm B}dv}{f}, 
\end{equation}
where $h$ is the Planck constant, 
$S \mu^{2}$ is the line strength multiplied by the square 
of dipole moment, 
$R$  is the relative intensity of the main hyperfine transition 
with respect to the rest of hyperfine transitions, 
$g_{\rm u}$  is the statistical weight of the upper level, 
$T_{\rm ex}$ is the excitation temperature, 
$T_{\rm bg}$ is the back ground temperature, 
$E_{\rm u}$  is the energy of the upper state, 
$\nu$  is the transition frequency, 
$T_{\rm B}$ is brightness temperature,  
$\int T_{\rm B}dv$ is the velocity-integrated intensity, 
$f$ is the filling factor, 
and $Q_{\rm rot}$ is the partition function. 
Here $f$ is assumed to be 1 and $T_{\rm NH_3}$ 
approximates $T_{\rm ex}$ of molecular lines.

The molecular column density $N_{\rm H_{2}}$, 
gas mass  $M_{\rm gas}$, and volume density 
$n_{\rm H_{2}}$ are derived from the continuum 
emission with 
\begin{equation}
\label{N_H2}
N_{\rm H_{2}} = \eta \frac{S_{\nu}}{B_{\nu}(T) \, \kappa_{\nu} \, \Omega \, \mu \, m_{\rm H}},
\end{equation}

\begin{equation}
\label{Mgas}
M_{\rm gas} = \eta \frac{F_{\nu} \, \rm{D^2}}{\kappa_{\nu} \, B_{\nu}(T)},
\end{equation}

\begin{equation}
\label{n_H2}
n_{\rm H_{2}} =  \frac{M_{\rm gas}}{\frac{4}{3} \pi r^{3} \, \mu \, m_{\rm H}},
\end{equation}
where $\eta$=100 is the gas-to-dust ratio, $S_{\nu}$ is the peak 
flux density, 
$F_{\nu}$ is the measured integrated source flux,  
$\Omega$ is the beam solid angle, $m_{\rm H}$ is 
the mass of an hydrogen atom, $\mu$=2.8 is the mean molecular 
weight of the interstellar medium \citep{2008A&A...487..993K},  
$\kappa_{\nu}$ is the dust opacity at a frequency of $\nu$,  
D is the distance to the source, and $r$ is the radius 
of dense cores. 
We adopted a value of 0.9 cm$^{-2}$ g$^{-1}$ for 
$\kappa_{\rm 1.3 mm}$, which corresponds to the opacity 
of thin ice mantles and a gas density of 10$^{6}$ cm$^{-3}$  
\citep{1994A&A...291..943O}.

The flux density is converted to brightness temperature 
following\footnote{\url{https://science.nrao.edu/facilities/vla/proposing/TBconv}}
\begin{equation}
T_{\rm B} = 1.222 \times 10^{3} \frac{I}{\nu^{2} \, \theta_{\rm maj} \, \theta_{\rm min}},
\end{equation}
where $T_{\rm B}$ is the brightness temperature in K, 
$I$ is the flux in mJy beam$^{-1}$, $\nu$ is the frequency 
in GHz, and $\theta_{\rm maj}$ and $\theta_{\rm min}$ are the 
half-power beam widths along the major and minor 
axes, respectively.


\end{document}